\documentclass[aps, prl, reprint, citeautoscript, superscriptaddress, showkeys, showpacs]{revtex4-1}   %
\usepackage{xr}
\usepackage[fleqn,tbtags]{amsmath}
\usepackage{amsfonts, amssymb, amsxtra, textcomp}
\usepackage[]{graphicx}
\usepackage{grffile}
\usepackage{bbm}
\usepackage{bm}
\usepackage{color}
\usepackage{graphicx}

\allowdisplaybreaks
\usepackage{pdfpages}

\makeatletter
\AtBeginDocument{\let\LS@rot\@undefined}
\makeatother

\usepackage{color}                                      
\definecolor{d_red}{cmyk}{0.00, 0.81, 1.00, 0.27}
\definecolor{d_orange}{cmyk}{0.00, 0.33, 1.00, 0.00}
\definecolor{d_blue}{cmyk}{0.78, 0.47, 0.00, 0.20}
\definecolor{d_lgreen}{cmyk}{0.07, 0.00, 0.79, 0.29}
\definecolor{d_green}{cmyk}{0.66, 0.00, 0.71, 0.56}
\definecolor{d_blue}{cmyk}{0.78, 0.47, 0.00, 0.20}
\definecolor{d_dblue}{cmyk}{0.91, 0.79, 0.00, 0.22}
\definecolor{d_pink}{cmyk}{0.0, 0.79, 0.37, 0.29}
\definecolor{d_purple}{cmyk}{0.16, 0.54, 0.00, 0.70}
\definecolor{d_paleblue}{cmyk}{0.669, 0.338, 0.00, 0.373}
\definecolor{d_dpaleblue}{cmyk}{0.441, 0.290, 0.00, 0.580}
\definecolor{d_brown}{cmyk}{0.0, 0.490, 0.930, 0.350}
\definecolor{d_turquoise}{cmyk}{0.630, 0.04, 0.0, 0.440}

\newcommand{\av}[1]{\langle #1 \rangle}

\newcommand{\bfmm}{{\boldsymbol{M}}}

\newcommand{\bfrr}{{\boldsymbol{R}}}
\newcommand{\bfss}{{\boldsymbol{S}}}

\newcommand{\bfk}{{\boldsymbol{k}}}

\newcommand{\bfp}{{\boldsymbol{p}}}
\newcommand{\bfq}{{\boldsymbol{q}}}

\newcommand{\bfs}{{\boldsymbol{s}}}

\newcommand{\bfx}{{\boldsymbol{x}}}
\newcommand{\bfy}{{\boldsymbol{y}}}

\def\bmx{\begin{pmatrix}}
\def\emx{\end{pmatrix}}

\usepackage{hyperref}

\begin{document}

\title{Enhanced nematic fluctuations near an antiferromagnetic Mott insulator and
possible application to high-$T_{c}$ cuprates }

\author{Peter P. Orth}

\affiliation{Department of Physics and Astronomy, Iowa State University, Ames,
Iowa 50011, USA}

\author{Bhilahari Jeevanesan}

\affiliation{Institute for Theory of Condensed Matter, Karlsruhe Institute of
Technology (KIT), 76131 Karlsruhe, Germany}

\author{Rafael M. Fernandes }

\affiliation{School of Physics and Astronomy, University of Minnesota, Minneapolis,
Minnesota 55455, USA}

\author{J\"org Schmalian}

\affiliation{Institute for Theory of Condensed Matter, Karlsruhe Institute of
Technology (KIT), 76131 Karlsruhe, Germany}

\affiliation{Institute for Solid State Physics, Karlsruhe Institute of Technology
(KIT), 76131 Karlsruhe, Germany}
\begin{abstract}
Motivated by the widespread experimental observations of nematicity in strongly underdoped cuprate superconductors, we investigate the possibility of enhanced nematic fluctuations in the vicinity of a Mott insulator that displays N\'eel-type antiferromagnetic order.  By performing a strong-coupling expansion of an effective model that contains both Cu-$d$ and O-$p$ orbitals on the square lattice, we demonstrate that quadrupolar fluctuations
in the $p$-orbitals inevitably generate a biquadratic coupling between the spins of the $d$-orbitals. The key point revealed by our classical Monte Carlo simulations and large-$N$ calculations is that the biquadratic term favors local stripe-like magnetic fluctuations, which result in an enhanced nematic susceptibility that onsets at a temperature scale determined by the effective Heisenberg exchange $J$. We discuss the impact of this type of nematic order on the magnetic spectrum and outline possible implications on our understanding of nematicity in the cuprates.
\end{abstract}
\maketitle
\section*{Introduction}
\begin{figure}[t]
\includegraphics[width=\linewidth]{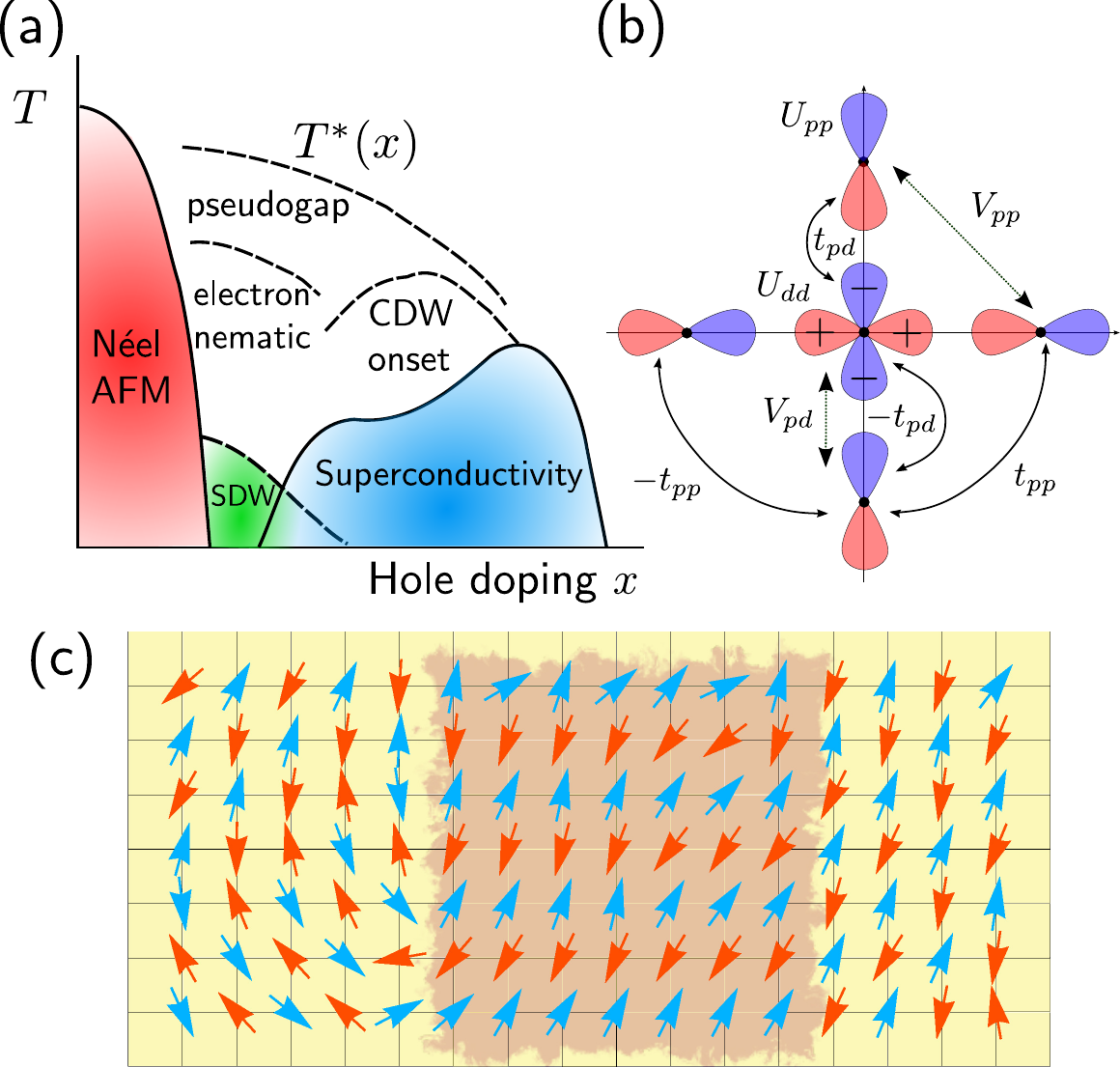} \caption{\textbf{Microscopic manifestation of nematic fluctuations.} (a) Schematic
phase diagram of hole-doped cuprate superconductors. (b) The microscopic
model we use contains a Cu $3d_{x^{2}-y^{2}}$ orbital (center orbital)
and O $2p_{x}$ and $2p_{y}$ orbitals in a single unit cell. The
hopping parameters are given by $t_{pd},t_{pp}$ and interactions
are considered on-site $U_{dd},U_{pp}$ and between nearest-neighbors
$V_{pd},V_{pp}$. (c) Simple illustration that nematic fluctuations
induce a short-ranged magnetic stripe ordered region (light red) within
a N\'eel ordered background (yellow). This snapshot is taken from our classical
Monte Carlo simulations. Red and blue color of the arrows denote out-of-the-plane
components of the spins. }
\label{fig:1} 
\end{figure}

Hole-doped cuprates are susceptible to a variety of different types
of electronic order in the underdoped regime. Examples include tendencies
towards charge order~\cite{Wu-Nature-2011,Ghiringhelli-2012,Chang2012,LeBoeuf2012},
which becomes long-ranged in the presence of large magnetic fields~\cite{Wu-Nature-2011,LeBoeuf2012},
and tendencies towards nematic order~\cite{Ando2002,Hinkov-Science-2008,Daou-Nature-2010,Lawler-Nature-2010,Cyr-PRB-2015,Ramshaw-npj-2017},
characterized by the breaking of the tetragonal symmetry of the system~\cite{Kivelson-RMP-2003,Vojta-AdvPhys-2009}.
The fact that these tendencies appear in the region of the phase diagram
where a pseudogap is also observed (see schematic Fig.~\ref{fig:1})
suggest a close interplay between these seemingly different phenomena,
a topic that remains widely debated in the field (for recent reviews,
see \cite{KeimerKivelson-Nature-2015,Fradkin2015}).

Although the microscopic mechanisms behind these different ordering
tendencies, and particularly of nematicity, remain unsettled, they
have been the subject of many different theoretical proposals (see,
for instance, \cite{Kivelson-Nature-98,Yamase-JPhysSocJpn-2000,Kivelson-PRB-2004,Yamase-PRB-2006,Yamase-PRB-2009,Okamoto-PRB-2010,Fischer-PRB-2011,Andersen-EPL-2012,Bulut-PRB-2013,Fischer-NJP-2014,Volkov-PRB-2016,Yamase-PRB-2009,Chubukov-PRB-2014,Schuett-PRL-2015,Nie-PRB-2017,Scheurer-PRL-17,Tsuchiizu-PRB-2018},
and also the reviews~\cite{Kivelson-RMP-2003,Vojta-AdvPhys-2009}). While
a complete theory for nematicity in the cuprates is beyond the scope
of our work, here we show that an important contribution to the nematic
susceptibility arises already near the Mott (or more precisely, charge-transfer~\cite{Zaanen-PRL-1985})
insulating state of the parent compound. For the rest of the paper,
thus, we focus only on the spin correlations near the Mott state,
and neglect other phenomena that are certainly important for a complete
description of the hole-doped cuprates, and which may also be important
to describe nematicity, such as charge order, pseudogap, time-reversal
symmetry-breaking, pair-density waves, and superconductivity \cite{KeimerKivelson-Nature-2015,Fradkin2015}.

To be more specific, we consider the so-called Emery model \cite{Emery-PRL-1987},
an effective model that attempts to capture both Cu and O low-energy
degrees of freedom by introducing $d_{x^{2}-y^{2}}$ orbitals on the
sites of the square lattice and $p_{x}$ ($p_{y}$) orbitals on the
horizontal (vertical) bonds. Consider first the case where only $d$-orbitals
are present. In the half-filled Mott insulating state, the charge
degrees of freedom are quenched, and the low-energy physics is described
completely in terms of an AFM Heisenberg interaction $J$ between
the $d$-orbital spins, which ultimately gives rise to a N\'eel AFM
ground state. Upon light hole-doping, the effective Hamiltonian is
known as the $t-J$ model \cite{Lee-RMP-2006}:

\begin{equation}
H_{t-J}=\sum_{ij\alpha}t_{ij}\tilde{d}_{i\alpha}^{\dagger}\tilde{d}_{j\alpha}+J\sum_{\left\langle ij\right\rangle }\left(\mathbf{S}_{i}\cdot\mathbf{S}_{j}-\frac{1}{4}n_{i}n_{j}\right).\label{H_tJ}
\end{equation}

Here, $t_{ij}$ denotes the hole hopping parameters and $J$ the AFM
exchange coupling. The operator $\mathbf{S}_{i}=\frac{1}{2}\sum_{\alpha\beta}\tilde{d}_{i\alpha}^{\dagger}\boldsymbol{\sigma}_{\alpha\beta}\tilde{d}_{i\beta}$
describes the $d$-orbital spin and $n_{i}=\sum_{\alpha}\tilde{d}_{i\alpha}^{\dagger}\tilde{d}_{i\alpha}$
the corresponding charge. The strong local Coulomb interaction is
incorporated in terms of the hole creation operator $\tilde{d}_{i\alpha}^{\dagger}=\left(1-n_{i\bar{\alpha}}\right)d_{i\alpha}^{\dagger}$,
reflecting the fact that double occupancy of the sites is not allowed
near the Mott insulating state.

As we demonstrate below via a strong coupling expansion of the Emery
model, the inclusion of the $p$-orbitals leads to an important additional
term in the $t-J$ Hamiltonian. While the two terms in Eq. (\ref{H_tJ})
remain the same, albeit with a different microscopic expression for
$J$, non-critical quadrupolar fluctuations of the $p$-orbitals,
enhanced by the repulsion between $p$-orbitals, generate a positive
biquadratic coupling $K>0$ between the $d$-orbital spins:

\begin{equation}
H_{K}=-K\sum_{i}\left[\mathbf{S}_{i}\cdot\left(\mathbf{S}_{i-\hat{x}}+\mathbf{S}_{i+\hat{x}}-\mathbf{S}_{i-\hat{y}}-\mathbf{S}_{i+\hat{y}}\right)\right]^{2},
\label{eq:HtJK}
\end{equation}
resulting in an effective $t-J-K$ Hamiltonian, $H_{t-J-K}=H_{t-J}+H_{K}$. 

Using classical Monte Carlo and large-$N$ analytical methods, we find that
the main consequence of $H_{K}$ is to enhance the static electronic
nematic susceptibility $\chi_{\mathrm{nem}}$ near the AFM-Mott insulating
state. However, $\chi_{\mathrm{nem}}$ is not found to diverge on
its own \textendash{} instead, it peaks at a temperature
scale proportional to $J$, instead of $K$. The location of the peak depends on the relative strength of quantum and thermal fluctuations and shifts towards smaller temperatures for larger quantum fluctuations. As illustrated in Fig.~\ref{fig:1}(c),
the enhancement of nematic fluctuations promoted by $H_{K}$ has its
origins on the \emph{short-ranged} magnetic stripe ordered regions
that this term favors within the (much longer ranged) N\'eel ordered background. Consequently,
within the $t-J-K$ model, the onset of nematic order requires an
additional symmetry breaking field that can take advantage of the
enhanced susceptibility. While a more detailed discussion of the application
of these results to the cuprates is left to the end of this paper,
we note that this mechanism for enhanced nematic susceptibility can
in principle be combined with other mechanisms proposed in the literature
to yield long-range nematic order. Detailed reviews on the proposed
mechanisms for nematicity in cuprates, both within weak and strong
coupling regimes, can be found for instance in Refs.~\cite{Kivelson-RMP-2003,Vojta-AdvPhys-2009}.

\section*{Results}

\emph{Microscopic model.} Our starting point is the interacting three-orbital
Emery model $H=H_{0}+H_{U}+H_{V}$~\cite{Emery-PRL-1987}. As depicted
in Fig.~\ref{fig:1}(b), it includes the $d_{x^{2}-y^{2}}$ Cu orbital
with creation operator $d_{i,\sigma}^{\dag}$ at Bravais lattice position
$\bfrr_{i}$ and spin $\sigma$ as well as the $p_{x}$ and $p_{y}$
O orbitals with creation operators $p_{i+\frac{\hat{x}}{2},\sigma}^{\dag}$
and $p_{i+\frac{\hat{y}}{2},\sigma}^{\dag}$. The non-interacting
part $H_{0}$ includes hopping between $p$-orbitals with (amplitude
$t_{pp}$) and between $d$- and $p$-orbitals (with amplitude $t_{pd}$).
The corresponding sign factors of the hopping elements follow from
the phases of the orbitals (see Fig.~\ref{fig:1}(b))~\cite{Emery-PRL-1987}.
In addition, $H_{0}$ contains on-site terms where the energy difference
between Cu and O orbitals is given as $\Delta=\varepsilon_{p}-\varepsilon_{d}$.
Interactions are included on-site $H_{U}=U_{dd}\sum_{i}n_{i,\uparrow}^{d}n_{i,\downarrow}^{d}+\frac{U_{pp}}{2}\sum_{i,u}n_{i+u,\uparrow}^{p}n_{i+u,\downarrow}^{p}$
with $u\in\{\frac{\hat{x}}{2},\frac{\hat{y}}{2}\}$, and number operators
$n_{i,\sigma}^{d}=d_{i,\sigma}^{\dag}d_{i,\sigma}$ and $n_{i+u,\sigma}^{p}=p_{i+u,\sigma}^{\dag}p_{i+u,\sigma}$.
We also consider nearest-neighbor interactions $H_{V}=\frac{V_{pp}}{2}\sum_{i,u,u'}n_{i+u}^{p}n_{i+u+u'}^{p}+V_{pd}\sum_{i,u}n_{i}^{d}n_{i+u}^{p}$
with $u'\in\{\pm\frac{1}{2}(\hat{x}+\hat{y}),\pm\frac{1}{2}(\hat{x}-\hat{y})\}$
and $n_{i}^{d}=\sum_{\sigma}n_{i,\sigma}^{d}$, $n_{i+u}^{p}=\sum_{\sigma}n_{i+u,\sigma}^{p}$.

The largest energy scales are the local repulsion $U_{dd}$ between
$d$-orbitals and the charge-transfer energy $\Delta$ (with $U_{dd}$
much larger than $\Delta$), suggesting a strong coupling expansion
in small $t_{ij}\ll\Delta,U_{dd}-\Delta$. This yields a description
in terms of localized $d$-orbital spins $\bfss_{i}$ coupled to mobile
$p$-orbital holes. An expansion up to fourth order in the hopping
term $t_{pd}$ was performed in Ref.~\cite{Zaanen-PRB-1988,Kolley-JPhysC-1992}. There appear
Kondo-like exchange couplings $\propto\bfss_{i}\cdot\bfs_{i+u_{1},i+u_{2}}$
between the $d$- and $p$-orbital spin-densities $\bfs_{i+u_{1},i+u_{2}}=\frac{1}{2}\sum_{\tau,\tau'}p_{i+u_{1},\tau}^{\dagger}\boldsymbol{\sigma}_{\tau\tau'}p_{i+u_{2},\tau'}$~\cite{ZhangRice-PRB-1988},
the familiar Heisenberg spin exchange term $J\sum_{\langle i,j\rangle}\bfss_{i}\cdot\bfss_{j}$
and terms that renormalize the $p$-orbital hole dispersion. For details
we refer to the Methods section and the Supplementary Information. The Kondo-like terms also modify the hole
dispersion as the tunneling process of holes through a $d$-orbital
becomes spin dependent. For example, tunneling through a background
of N\'eel ordered $d$-orbital spins leads to the spin dependent hopping
parameters $t_{a}=\frac{t_{pd}^{2}}{2}(\frac{1}{\Delta}+\frac{3}{U_{dd}-\Delta})$
and $t_{b}=\frac{t_{pd}^{2}}{2}(\frac{3}{\Delta}+\frac{1}{U_{dd}-\Delta})$
for holes with spin parallel and antiparallel to the central $d$-orbital
spin~\cite{Fischer-NJP-2014}. The hole Fermi surface thus appears at momenta $\bfk=(\pm\frac{\pi}{2},\pm\frac{\pi}{2})$
for small doping $n_{p}$.

\begin{figure}[t!]
\includegraphics[width=1\linewidth]{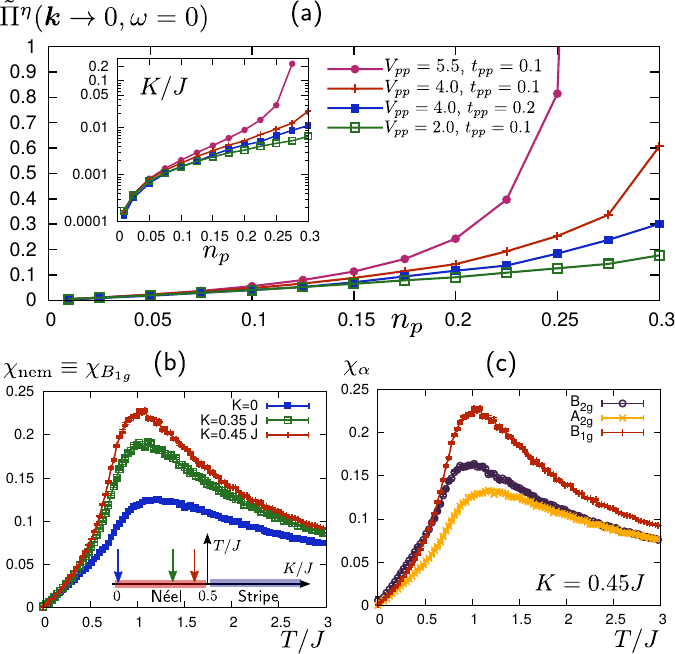} \caption[]{ \textbf{Strength of biquadratic exchange K within Emery model and
enhanced nematic spin fluctuations.} (a) Renormalized quadrupolar
oxygen density response function $\tilde{\Pi}_{\bfk=0}^{\eta}=\frac{1}{2}[(\Pi_{\bfk=0}^{\eta})^{-1}-U_{\bfk=0}]^{-1}$
as a function of $p$-orbital holes $n_{p}$ (per planar $d$-orbital)
obtained within the three-band Emery model at low temperature $T=10^{-2}t_{pp}$
and fixed $n_{d}=1$. The interaction of the mobile holes with the
antiferromagnetic N\'eel background of $d$-orbital spins is fully taken
into account. Other parameters are set to $t_{pd}=1$, $\Delta=2.5$,
$U_{dd}=11$ and $V_{pd}=V_{pp}$ such that $U_{\bfk=0}=V_{pp}-\frac{U_{pp}}{8}$.
We use $U_{pp}=5.5$ for $V_{pp}=5.5$ and $U_{pp}=4.5$ for all other
values of $V_{pp}$. Amplitude of oxygen quadrupolar fluctuations
increases with $V_{pp}$ and smaller oxygen bandwidth, e.g., smaller
$t_{pp}$. The inset shows the resulting value of $K/J\propto(J'^{2}/J)(n_{p}/t_{pp})$
(at small $n_{p}$) from which we conclude that an enhancement of
the fluctuations by $V_{pp}$ is crucial for a significant biquadratic
exchange coupling. Note that we have approximated $\Pi_{\bfk}^{\eta}\approx\Pi_{\bfk=0}^{\eta}$
for simplicity, which does not affect our conclusion. Panels (b-c)
show the static nematic susceptibility $\chi_{\text{nem}}$ in Eq.~(\ref{eq:chi)nem})
for the $H_{t-J-K}$ model of Eq.~(\ref{eq:HtJK}) at half-filling
as a function of temperature $T$ obtained by Monte Carlo simulations
of classical spins. A non-zero $K$ enhances the response in the nematic
$B_{1g}$ channel only. Inset phase diagram shows that we are investigating
$\chi_{\text{nem}}$ above the N\'eel ordered state. For consistency
with the known spin-wave spectrum, we consider a small ferromagnetic
next-nearest-neighbor exchange $J_{2}=-0.1J$. }
\label{fig:2} 
\end{figure}


Most notably for our considerations, the strong coupling expansion
also yields a spin exchange term that depends on the occupation of
the intermediate $p$-orbital between $d$-orbital sites: 
\begin{equation}
H_{J'}=-J'\sum_{i,\delta}n_{i+\frac{\delta}{2}}^{p}\bfss_{i}\cdot\bfss_{i+\delta}
\label{eq:2}
\end{equation}
where $\delta\in\{\pm\hat{x},\pm\hat{y}\}$ and the spin exchange
coupling constant is given by $J'=\sum_{n=0}^{3}\frac{t_{pd}^{4}{\rm sign}\left(3-2n\right)}{\Delta^{3-n}\left(U_{dd}-\Delta\right)^{n}}$.
Note that $J=\sum_{n=0}^{2}\frac{t_{pd}^{4}(4-n^{2}-\delta_{n,2})}{2\Delta^{3-n}(U_{dd}-\Delta)^{n}}$
and in the large-$U_{dd}$ limit both are of the same order $J'/J\rightarrow1/2$. Oxygen charge fluctuations
thus not only renormalize the Heisenberg exchange via the Kondo coupling
terms, but, as we show now, also lead to the biquadratic spin exchange
interaction $K$ in Eq.~\eqref{eq:HtJK}.

We derive the biquadratic exchange $K$ by first decomposing the $p$-orbital
densities as $n_{i+\frac{\hat{x}}{2}}^{p}=n_{i}^{p}+\eta_{i}$ and
$n_{i+\frac{\hat{y}}{2}}^{p}=n_{i}^{p}-\eta_{i}$, where $\eta_{i}$
is the quadrupolar (nematic) component of the oxygen charge density~\cite{Fischer-PRB-2011}.
The combination of on-site and nearest-neighbor Coulomb interactions
between $p$-orbital leads to a term $(-2\sum U_{\bfk}\eta_{\bfk}\eta_{-\bfk})$
in the Hamiltonian, where $U_{\bfk}=\frac{1}{4}(V_{pp}\text{Re}f_{\bfk}-\frac{U_{pp}}{2})$
and $f_{\bfk}=1+e^{-ik_{x}}+e^{ik_{y}}+e^{i(k_{y}-k_{x})}$. Integrating
out the quadrupolar charge fluctuations associated with the $p$-orbitals
(details of this analysis are presented in the Methods section and the Supplementary Information) yields
the result for the biquadratic exchange interaction in Eq.~\eqref{eq:HtJK}
with:
\begin{equation}
K=\frac{J'^{2}}{2}\frac{\int_{\mathbf{k}}\Pi_{\mathbf{k}}^{\eta}}{1-(V_{pp}-\frac{U_{pp}}{8})\Pi_{\mathbf{k}=0}^{\eta}}>0\,.\label{eq:5}
\end{equation}
Here, $\Pi_{\mathbf{k}}^{\eta}=-\int_{\mathbf{q},\omega'}\text{Tr}[G_{\mathbf{\mathbf{q},\omega'}}^{p}(\tau^{z}\sigma^{0}s^{0})G_{\mathbf{q}+\mathbf{k},\omega'}(\tau^{z}\sigma^{0}s^{0})]$
is the bare $p$-orbital charge susceptibility in the quadrupolar
(i.e. nematic) channel We have used the long-wavelength approximation
in the denominator for simplicity (the full expression can be found
in the Supplementary Information) and yields qualitatively identical results).
The Pauli matrices $\tau^{j}$, $s^{j}$ and $\sigma^{i}$ act in
orbital $(p_{x},p_{y})$, spin and reduced wavevector $\bigl(\bfk,\bfk+(\pi,\pi)\bigr)$
space, respectively. Note that the presence of the AF background of
$d$-orbital spins leads to a doubling of the unit cell, and thus
$\int_{\bfq}\equiv\int_{\text{mBZ}}\frac{d^{2}q}{2\pi^{2}}$ is an
integration over the magnetic Brillouin zone (mBZ). Explicit expressions
for the $p$-orbital Green's functions $G_{\mathbf{q},\omega}^{p}$
are given in the Methods section and the Supplementary Information and yield 
\begin{align}
\Pi_{\bfk}^{\eta}=-\int_{\bfq}\sum_{i,j=1}^{4}\frac{v_{ij;\bfq,\bfq+\bfk}\{n_{F}(\xi_{i,\bfq})-n_{F}(\xi_{j,\bfq+\bfk})\}}{\epsilon_{i,\bfq}-\epsilon_{j,\bfq+\bfk}}\label{eq:8}
\end{align}
with $\epsilon_{i,\bfq}$ being the renormalized $p$-orbital dispersion
and $\xi_{i,\bfq}=\epsilon_{i,\bfq}-\mu$ with the chemical potential
$\mu$. The matrix elements $v_{ij;\bfq,\bfq+\bfk}=U_{\bfq}^{\dag}(\tau^{z}\sigma^{0})U_{\bfq-\bfk}$
contain $U_{\bfq}$, which are unitary matrices that transform between
orbital/reduced $\bfk$-space ($\sigma^{i}\otimes\tau^{j})$ and band
space. Most importantly, Eq.~\eqref{eq:5} makes it clear that the
biquadratic exchange $K$ in Eq.~\eqref{eq:HtJK} is a direct consequence
of quadrupolar oxygen charge fluctuations. These are a generic feature
of the model and exist even if the $p$-orbital holes are not dressed
by $d$-orbital spins.

The oxygen quadrupolar susceptibility $\Pi_{\bfk}^{\eta}$ (and thus
$K$) is strictly positive for all $\bfk$ and is determined by the
occupation number difference between the different oxygen bands. In
the relevant regime of small hole fillings $n_{p}\ll1$, the response
approaches a value $\Pi_{\bfk=0}^{\eta}\propto n_{p}$ at low $T$,
peaks around $T\approx|\mu|$ and vanishes as $1/T$ at large $T$.
The response increases for smaller bandwidth, e.g., smaller $t_{pp}$.
This is derived explicitly in the Supplementary Information for a simpler two-band model
that neglects the interaction with the AF background. It also holds
true numerically for the full four-band model, as shown in Fig.~\ref{fig:2}(a),
where we present results for the renormalized quadrupolar response
$\tilde{\Pi}_{\boldsymbol{k}=0}^{\eta}=\frac{1}{2}\bigl[(\Pi_{\mathbf{k}=0}^{\eta}\bigr)^{-1}-U_{\bfk=0}\bigr]^{-1}$
and for the resulting $K/J$ within the microscopic four-band model.
In the calculation we keep $n_{d}=1$, assuming that holes are doped
into the $p$-orbitals, but we take the interaction of the mobile
holes with the AF background of $d$-orbital spins fully into account.
We clearly observe that a large nearest-neighbor repulsion $V_{pp}$
and a small bandwidth $t_{pp}$ enhance $K$ (see Eq.~\eqref{eq:5}).
Our results also indicate that an enhancement of the quadrupolar density
fluctuations by $V_{pp}$ is necessary for a significant biquadratic
exchange coupling. This follows from $K/J\propto(J'^{2}/J)(n_{p}/t_{pp})$
at small $n_{p}$ where $J'=0.08$, $J=0.15$ for the parameters in
Fig.~\ref{fig:2}. Finally, while phonon modes in the same channel
are, by symmetry, allowed to give rise to similar behavior, the electronic
mechanism for biquadratic exchange is expected to be quantitatively
much stronger.

\emph{Enhanced nematic susceptibility.} The implications of $H_{K}$
can be better understood in the limit of $K>J$. In this case, the
AFM ground state is no longer the N\'eel configuration with ordering
vector $\mathbf{Q}=\left(\pi,\pi\right)$, but the striped configuration
with $\mathbf{Q}=\left(\pi,0\right)$ or $(0,\pi)$. While the limit
of large $K/J$ is clearly not realized in the cuprates, it reveals
that $H_{K}$ supports quantum and classical fluctuations with local
striped-magnetic order that have significant statistical weight.

We qualitatively demonstrate this behavior in Fig.~\ref{fig:1}(c)
by showing typical spin configurations of a Monte Carlo analysis of
$H_{t-J-K}$ in the limit of classical spins and where the kinetic
energy of the holes is ignored. One clearly sees local striped-magnetic
fluctuations (light red background) in an environment of N\'eel ordered
spins (yellow background). Configurations with parallel spins along
the $x$-axis and along the $y$-axis occur with equal probability,
hence preserving the tetragonal symmetry of the system. If one, however,
weakly disturbs tetragonal symmetry, e.g. by straining one of the
axes, this balance is disturbed and one favors striped configurations
of one type over the other. 

The behavior described above can be quantified in terms of the composite
spin variable: 
\begin{equation}
\varphi_{i}=\mathbf{S}_{i}\cdot\left(\mathbf{S}_{i-\hat{x}}+\mathbf{S}_{i+\hat{x}}-\mathbf{S}_{i-\hat{y}}-\mathbf{S}_{i+\hat{y}}\right)\label{eq:phi_Ising}
\end{equation}
which changes sign under a rotation by $\pi/2$. Note that the square
of this term, which appears in Eq.(\ref{eq:HtJK}), is invariant under
this transformation, and therefore is fully consistent with the four-fold
symmetry of the Emery model. While $\left\langle \varphi_{i}\right\rangle =0$
for realistic values of $K$ (and in the absence of external strain),
the static nematic susceptibility 
\begin{equation}
\chi_{{\rm nem}}\left(T\right)=\int_{0}^{1/T}d\tau\sum_{i}\left\langle \mathcal{T}_{\tau}\varphi_{i}\left(\tau\right)\varphi_{0}\left(0\right)\right\rangle 
\label{eq:chi)nem}
\end{equation}
is a measure for the increased relevance of local stripe magnetic
configurations. Here, $\mathcal{T}_{\tau}$ denotes imaginary time
ordering.

We present a quantitative demonstration that the biquadratic exchange
$K$ yields an enhanced nematic susceptibility in the $B_{1g}$ ($x^{2}-y^{2}$)
symmetry channel in Fig.~\ref{fig:2}(b-c). It contains Monte Carlo
results for $\chi_{{\rm nem}}$ for a collection of classical Heisenberg
spins that interact according to the $H_{J-K}$ model. For consistency
with the known spin-wave spectrum, we have included an additional
small second-neighbor exchange $J_{2}=-0.1J$ in the simulation. One
clearly sees that the biquadratic term $K$ enhances the nematic response
in the $B_{1g}$ channel, corresponding to an inequivalence between
the $x$ and $y$ axes. In the limit of classical spins, the nematic
susceptibility $\chi_{{\rm nem}}\left(T\right)$ is non-monotonic,
peaking at a temperature governed by the effective exchange interaction
of the spins, $T_{{\rm nem}}\sim J$, which is independent on $K$.

\begin{figure}[h!]
\includegraphics[width=1\linewidth]{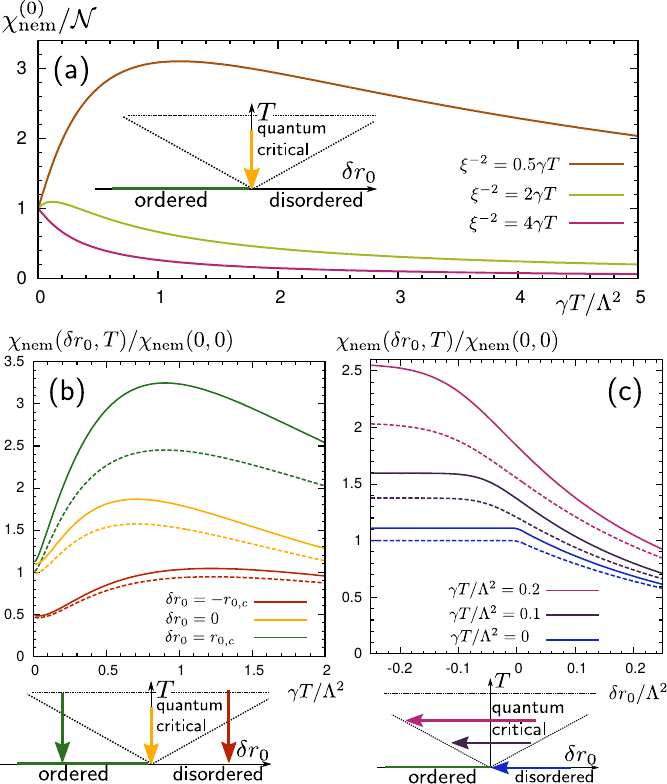} \caption{\textbf{Nematic susceptibility including quantum fluctuations.} Effects
of quantum fluctuations are included in the analytical treatment of
a soft-spin field theory of the $H_{J-K}$ model. Panel (a) shows
the bare nematic susceptibility $\chi_{\text{nem}}^{(0)}(\xi,T)$
(normalized to its $T=0$ value) as a function of temperature $T$,
right above the N\'eel QCP $r_{0}=r_{0,c}$ (see inset). Different curves
correspond to different functional behavior of the N\'eel correlation
length $\xi(T)^{-2}=a\gamma T$. The non-universal constant $a$ determines
whether $\chi_{\text{nem}}^{(0)}$ peaks at $T=0$ ($a>\pi$) or at
finite $T$ ($a<\pi$). Thus, $\chi_{\text{nem}}^{(0)}$ can exhibit
different shapes in different systems, as expected for a non-universal
susceptibility, which remains finite at the AFM QCP. Panels (b) and
(c) contain results of a large-$N$ treatment of the model, which
allows an explicit solution of $\xi(T,\delta r_{0})$. Dashed (solid)
lines are for $g=0$ ($g=0.1$), where $g\propto K/J$. Panel (b)
is for fixed distances to the QCP $\delta r_{0}=\{-r_{0,c},0,r_{0,c}\}$
(green, yellow, red; as indicated in the sketch below) and varying
$T$. Quantum fluctuations render the susceptibility at $T=0$ finite,
but have no strong effect on the finite temperature behavior. Importantly,
non-zero biquadratic exchange $g>0$ enhances the finite temperature
nematic response and increases the maximal value of $\chi_{\text{nem}}$
around $T\sim\Lambda^{2}/\gamma\sim J$ (with momentum cutoff $\Lambda=10$
and frequency cutoff $\gamma\Lambda_{\omega}=100$). Panel (c) is
for fixed temperatures $\gamma T/\Lambda^{2}=\{0,0.1,0.2\}$ (blue,
purple, magenta) and varying $\delta r_{0}$. It demonstrates that
the nematic response increases with the magnetic correlation length,
as the system approaches the QCP. The quartic coefficient is set to
$u/\gamma=50$ in (b) and $u/\gamma=5$ in (c). }
\label{fig:3} 
\end{figure}


The Monte Carlo results also display that $\chi_{{\rm nem}}\left(T\rightarrow0\right)\rightarrow0$,
which is a consequence of the classical nature of the spins in the
simulations and a resulting absence of (thermal) fluctuations in the
zero temperature limit. Quantum fluctuations crucially modify this
behavior and lead to $\chi_{{\rm nem}}\left(T\rightarrow0\right)>0$.
This is demonstrated in Fig.~\ref{fig:3}, where we present results
of an analytical calculation of the nematic response $\chi_{\text{nem}}$
that includes the effect of quantum fluctuations within a soft-spin
field-theoretical version of the spin degrees of freedom in Eq.~(\ref{eq:HtJK}).
After decoupling the biquadratic exchange term $K$ in the nematic
channel and taking the long wavelength limit, which is appropriate
to study the low-energy excitations, we obtain the effective action:
\begin{align}
S & =S_{{\rm dyn}}+\int_{r}\left[\left(\nabla\mathbf{n}_{r}\right)^{2}-\varphi_{r}\left(\left(\partial_{x}\mathbf{n}_{r}\right)^{2}-\left(\partial_{y}\mathbf{n}_{r}\right)^{2}\right)\right]\nonumber \\
 & +\int_{r}\left[r_{0}\mathbf{n}_{r}^{2}+\frac{u}{2}\left(\mathbf{n}_{r}\cdot\mathbf{n}_{r}\right)^{2}+\frac{\varphi_{r}^{2}}{2g}-h_{r}\varphi_{r}\right]
 \label{action}
\end{align}
where $\mathbf{n}_{r}$ is the $N=3$ component staggered N\'eel order
parameter, as used in the non-linear sigma model of Refs.~\cite{ChakravartyHalperinNelson-PRB-1989,Chubukov-PRB-1994}.
The parameter $r_{0}$ controls the distance to the AFM N\'eel quantum
critical point located at $r_{0,c}$ For $\delta r_{0}\equiv r_{0}-r_{0,c}<0$,
the system has long-range AFM order at $T=0$, whereas for $\delta r_{0}>0$
it is in the paramagnetic phase (see sketches at the bottom of
Fig.~\ref{fig:3}), and the interaction parameters are $g\propto K/J>0$,
$u>g$. The integrations are over $\int_{r}\equiv\int_{0}^{1/T}d\tau\int d^{2}r$,
where $r=\left(\tau,\mathbf{r}\right)$ combines imaginary time $\tau$
and position $\mathbf{r}=(x,y)$. In addition, $\varphi_{r}$ is the
nematic order parameter of Eq.~(\ref{eq:phi_Ising}) and $h_{r}$
is an external strain field. The quantum dynamics of the N\'eel order
parameter is governed by $S_{{\rm dyn}}=\int_{q}f\left(\omega_{n}\right)\mathbf{n}_{q}\cdot\mathbf{n}_{-q}$,
where $f\left(\omega_{n}\right)\propto\omega_{n}^{2}$ at half filling,
while $f\left(\omega_{n}\right)=\gamma\left|\omega_{n}\right|$ was
proposed to describe particle-hole excitations, and will be used below
as we describe the system away from half-filling $n_{p}>0$. Here,
$q=\left(\omega_{n},\mathbf{q}\right)$ combines Matsubara frequency
$\omega_{n}$ and momentum $\mathbf{q}$ (measured relative to the
AFM ordering vector $\mathbf{Q}=\left(\pi,\pi\right)$) and $\int_{q}\equiv T\sum_{n}\int\frac{d^{2}q}{\left(2\pi\right)^{2}}$.

The nematic susceptibility in Eq.~\eqref{eq:chi)nem} can be obtained
for general $N$ and reads (see Methods section and Supplementary Information): 
\begin{equation}
\chi_{{\rm nem}}=\frac{\chi_{{\rm nem}}^{\left(0\right)}}{1-\frac{g}{N}\chi_{{\rm nem}}^{\left(0\right)}},\label{eq:chinemN}
\end{equation}
where the bare nematic susceptibility is given by $\chi_{{\rm nem}}^{\left(0\right)}=\frac{N}{2}\int_{q}\frac{|\boldsymbol{q}|^{4}\cos^{2}(2\theta)}{\left(\xi^{-2}+|\boldsymbol{q}|^{2}+f(\omega_{n})\right)^{2}}$
with $\bfq=|\boldsymbol{q}|(\cos\theta,\sin\theta)$. Here, $\xi$
is the magnetic correlation length for N\'eel order, that includes interaction
corrections and diverges at the AFM phase transition. Right above
the quantum critical point at $\delta r_{0}=0$, one finds $\xi^{-2}=a\gamma T$
with non-universal constant $a$. As shown in Fig.~\ref{fig:3}(a),
the exact shape of $\chi_{\text{nem}}^{(0)}(T)$ depends on this non-universal
parameter $a$, which depends, for example, on the interaction parameter
$u$ or the lattice constant. While $\chi_{\text{nem}}^{(0)}$ peaks
at finite temperatures for $a<\pi$, which is similar to the classical
case, the maximum occurs at $T=0$ for $a>\pi$. Note that nematic
correlations remain finite ranged at the AFM quantum critical point
and universal behavior of $\chi_{\text{nem}}^{(0)}$ is not guaranteed
(in contrast to the AFM susceptibility, which is universal). Being
a non-universal quantity, we thus expect that the precise shape of
$\chi_{\text{nem}}^{(0)}$ can be different for different systems.

In order to make analytic progress and calculate $a(u)$, or more
generally $\xi(T,\delta r_{0})$, we consider the limit of large $N$.
This approach led to important insights in both the description of
antiferromagnetic correlations of the cuprate parent compounds~\cite{Chubukov-PRB-1994}
and of nematic fluctuations of iron-based superconductors~\cite{Fernandes-PRB-2012}.
The magnetic correlation length $\xi$ is determined self-consistently
within large-$N$ for a given distance to the AFM quantum critical
point $\delta r_{0}$. Despite the similarity between Eq.~(\ref{eq:chinemN})
and the expression for the nematic susceptibility of iron-based superconductors~\cite{Fernandes-PRB-2012},
there are very important differences between the two systems. Because
the iron pnictides order magnetically in a striped configuration,
$\chi_{{\rm nem}}^{\left(0\right)}$ diverges when $\xi\rightarrow\infty$,
which guarantees that a nematic transition takes place already in
the paramagnetic state for any $g>0$. However, because our model
orders in a N\'eel configuration, $\chi_{{\rm nem}}^{\left(0\right)}$
remains finite even when $\xi\rightarrow\infty$. Although long-range
nematic order is not present, nematic fluctuations can be significantly
enhanced if the biquadratic exchange $K\propto g$ is sufficiently
large.

In Fig.~\ref{fig:3}(b,c) we show the nematic susceptibility obtained
within the large-$N$ approach (see Methods section and Supplementary Information). Like in the Monte Carlo
results (see Fig.~\ref{fig:2}), we observe, in Fig.~\ref{fig:3}(b),
a broad maximum at finite temperatures around $T\approx J$, corresponding
at $\delta r_{0}=0$ to $a<\pi$. The lattice cutoff $\Lambda$ plays
the role of $J$ in the continuum model. The effect of $g$, and thus
of the biquadratic exchange $K$, is to enhance the amplitude of the
peak (comparing dashed and solid lines). The pronounced peak of $\chi_{{\rm nem}}$
originates from the bare susceptibility $\chi_{{\rm nem}}^{\left(0\right)}$.
As discussed above, the bare response is in turn governed by the magnetic
correlation length $\xi$ that is set by $T/J$. Notably, at low temperatures,
quantum fluctuations render $\chi_{{\rm nem}}$ (and $\chi_{\text{nem}}^{(0)}$)
finite, in stark contrast to our MC results for classical spins. Keeping
$T$ fixed and varying the non-thermal tuning parameter $\delta r_{0}$,
we observe in Fig.~\ref{fig:3}(c) that the nematic response increases
for an increasing magnetic correlation length, i.e. N\'eel fluctuations
enhance the nematic susceptibility. This follows from the observation
that $\chi_{{\rm nem}}$$\left(\delta r_{0},T\right)$ is an increasing
function for decreasing $\delta r_{0}$.

\begin{figure}[t!]
\includegraphics[width=1\linewidth]{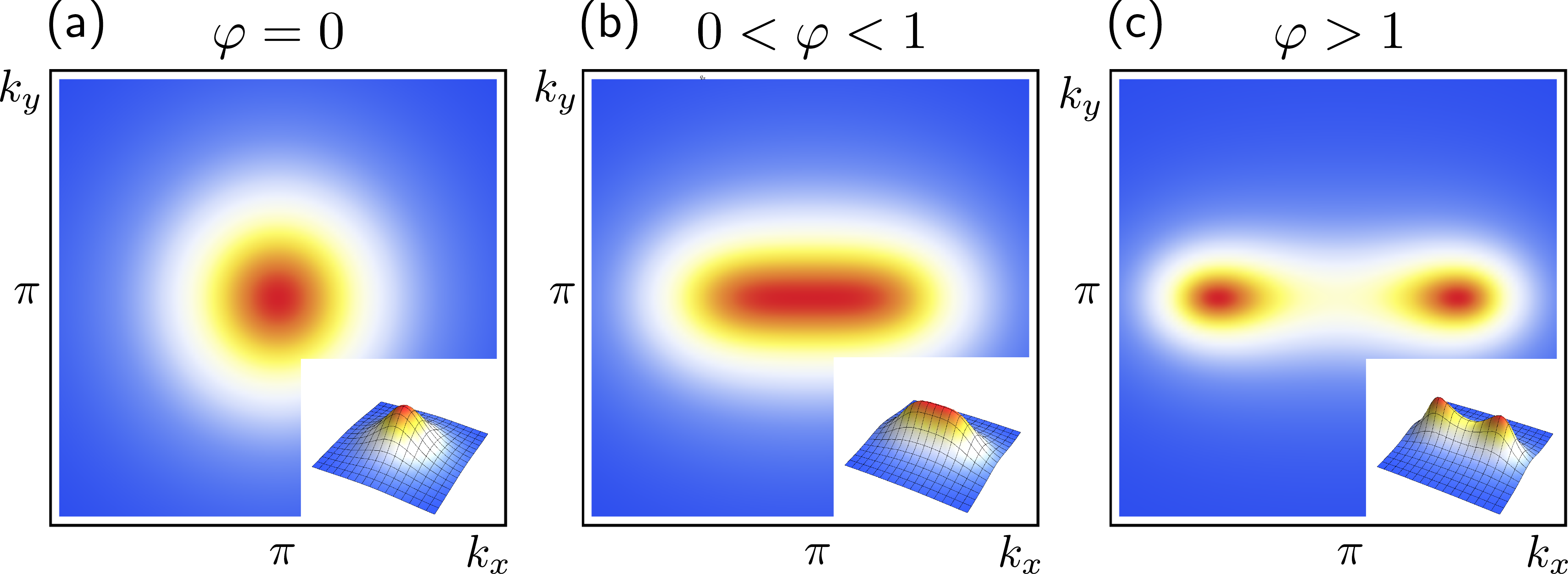} \caption{\textbf{Incommensurability transition induced by nematic order.} Panels
(a-c) schematically illustrate the effect of finite nematic order
$\varphi$ on the spin-spin correlation function. It shows $\chi_{\text{\text{AFM} }}(\boldsymbol{k})$
from Eq.~\eqref{chi_AFM} (including a fourth order term $\propto\boldsymbol{q}^{4}$)
for $\xi^{-2}=0.2$ and $\varphi=\{0.,1.,1.5\}$. In the absence of
nematic order ($\varphi=0$) the magnetic susceptibility peak is isotropic
around the N\'eel ordering vector $\boldsymbol{Q}=(\pi,\pi)$, but non-zero
$0<\varphi<1$ leads to an elliptic deformation of the peak. For larger
values of $\varphi>1$ the peak splits and two incommensurate scattering
peaks emerge at $(\pi\pm\delta,\pi)$. Note that, within our model,
the inequivalence of $x$ and $y$ direction appears only in response
to an external (or intrinsic) strain field that explicitly breaks
$C_{4}$ symmetry. }
\label{fig:4} 
\end{figure}

\emph{Consequences of long-range nematic order.} As we discussed above,
the nematic susceptibility does not diverge within our $H_{J-K}$
model. Nevertheless, it is interesting to study what happens to the
magnetic spectrum if nematic order is induced \textendash{} either
by the presence of a small tetragonal-symmetry breaking field $h$,
which can induce a sizable nematic order parameter $\varphi\approx\chi_{{\rm nem}}h$,
or by combination with other microscopic mechanisms for nematicity.
From the action in Eq.~(\ref{action}), we can readily obtain the
dynamic spin susceptibility in the presence of nematic order 
\begin{equation}
\chi_{{\rm AFM}}\left(\mathbf{Q}+\mathbf{q},\omega\right)=\frac{1}{\xi^{-2}+\mathbf{q}^{2}-\varphi\left(q_{x}^{2}-q_{y}^{2}\right)+f\left(\omega_{n}\right)},\label{chi_AFM}
\end{equation}
Therefore, as shown in Fig.~\ref{fig:4}, non-zero $\varphi$ modifies
the spin-spin structure factor near the N\'eel ordering vector $\mathbf{Q}$
from a circular shape, which preserves tetragonal symmetry, to an
elliptical shape, which breaks tetragonal symmetry. In addition, as
$\varphi$ increases, it shifts the maximum of $\chi_{{\rm AFM}}\left(\mathbf{Q}+\mathbf{q},\omega\right)$
from the commensurate $\mathbf{q}=0$ value to an incommensurate wavevector
$\mathbf{q}_{\mathrm{IC}}\neq0$, with $\mathbf{q}_{\mathrm{IC}}$
parallel to either the $x$ axis (if $\varphi>0$) or to the $y$
axis (if $\varphi<0$). Note that a somewhat related mechanism for
the incommensurate spin order, based on the $t-J$ model, was reported
in Refs.~\cite{Shraiman-PRL-1988,Sushkov-PRL-2005,Gabay-PhysicaC-1989}. Previous
works have also focused on nematicity arising from a pre-existing
incommensurability~\cite{Nie-PRB-2017}, whereas in our scenario incommensurate
magnetic order is a consequence of nematic order, caused by an enhanced
nematic susceptibility in the presence of an external symmetry breaking
field.

While these effects onset in the paramagnetic phase, the presence
of nematic order should also be manifested in the N\'eel ordered state
by the direction of the $d$-orbital moments, which would align parallel
to either the $x$ axis or to the $y$ axis~\cite{Nafradi-PRL-2016}.
Within our model, we argue that such an effect would arise in the
presence of spin-orbit coupling in the $p$-orbitals, which convert
an imbalance in the charge of the $p_{x}$ and $p_{y}$ orbitals into
a preferred direction for the $d$-orbital moment.

At first sight, one might anticipate that $K$ would affect the spin-wave
dispersion of the AFM N\'eel ground state~\cite{Coldea-PRL-2001}. As we show in the Supplementary Information,
however, the biquadratic exchange of Eq.~(\ref{eq:HtJK}) does not
modify the linearized classical spin-wave spectrum. The reason for
this peculiar behavior is that the biquadratic exchange annihilates
the classical N\'eel state, i.e. the vacuum of the linear spin wave
excitations: 
\begin{equation}
H_{K}\left|{\rm N\acute{e}el}\right\rangle =0\,.
\end{equation}

It is important to point out that all results discussed here were
obtained considering that the spins of the $H_{J-K}$ Hamiltonian
are treated as vectors, either classical or in the large-$N$ regime. It is interesting to ask what happens if one considers the quantum spin-$1/2$ case. It turns out that, for spin-$1/2$,
the biquadratic term $K>0$ transforms into an AFM next-nearest neighbor
bilinear exchange coupling, which certainly changes the spin-wave
spectrum. It remains an open question whether the results presented here remain unchanged if one performs this transformation from biquadratic to bilinear exchange in the microscopic model. Importantly, however, we note that a large AFM next-nearest neighbor exchange also favors a stripe magnetic state over a N\'eel state. Thus, the main ingredient
that enhances the nematic susceptibility in the classical spin case seems also to be present in the spin-$1/2$ case.

\section*{Discussion}

In summary, we showed via a strong coupling expansion of the Emery
model that quadrupolar charge fluctuations in the $p$-orbitals generate
a biquadratic exchange coupling between the $d$-orbital spins, extending
the celebrated $t-J$ model employed to describe lightly-doped Mott
insulators. The main effect of this biquadratic term is to enhance
$B_{1g}$ nematic fluctuations, which however is not translated into
a diverging nematic susceptibility. Importantly, the temperature at
which the nematic susceptibility peaks is set not by the biquadratic
coupling $K$, but by the standard nearest-neighbor exchange coupling
$J$. The peak position is controlled by the relative strength of
thermal versus quantum fluctuations, and moves from a temperature
of order $J$ for dominant thermal fluctuations towards zero for dominant
quantum fluctuations. The biquadratic exchange $K$, however, sets
the amplitude of the peak, and both increase for larger values of
the repulsion $V_{pp}$ between nearest-neighbor $p$-orbitals.

Thus, our main result is that magnetic correlations associated with
the Mott insulating state generate an enhanced nematic susceptibility, which is driven by quadrupolar oxygen density fluctuations. In the remainder of this section, we discuss the possible applications
of these results to the nematic tendencies observed in hole-doped
cuprates \cite{Kivelson-RMP-2003,Vojta-AdvPhys-2009}. The mechanism discussed
here does not lead to long-range nematic order on its own. However,
given the enhanced nematic susceptibility, it is expected that a small
tetragonal symmetry-breaking field would lead to a sizable nematic
order parameter. Such a symmetry-breaking field is naturally provided
by the CuO chains or double chains in YBa$_{2}$Cu$_{3}$O$_{7-\delta}$
and YBa$_{2}$Cu$_{4}$O$_{8}$, respectively. Interestingly, in YBCO,
several experimental observations are consistent with the existence
of an electronic nematic order parameter \cite{Ando2002,Hinkov-Science-2008,Daou-Nature-2010,Ramshaw-npj-2017}.
Whether the observed nematicity is the result of the intrinsic small
symmetry-breaking field combined with a large nematic susceptibility,
or the consequence of true long-range order that would onset even
if the chains were absent, remains to be determined.

Still in what concerns YBCO, it is interesting to note that nematic
order is observed already at rather small doping levels, in the vicinity
of the Mott insulating N\'eel state \cite{Hinkov-Science-2008,Haug-NJP-2010}. In this
region of the phase diagram, where our results are the most relevant,
the experimental nematic onset temperature is comparable to that of
the N\'eel transition temperature, which in turn is set by $J$. Of
course, since nematicity is not restricted only to the vicinity of
the N\'eel state, it is possible that there are different mechanisms
responsible for nematicity in different regions of the phase diagram
\cite{Nie-PRB-2017}.

Neutron scattering experiments in YBCO also reveal a strong feedback
of nematic order on the magnetic spectrum \cite{Hinkov-Science-2008,Haug-NJP-2010,Nafradi-PRL-2016}.
In particular, nematic order is manifested as an elliptical spin structure
factor centered at the N\'eel ordering vector. Upon lowering the temperature,
the peak splits and gives rise to two unidirectional incommensurate
peaks. These observations are qualitatively consistent with our results
for the effect of nematicity on the AFM magnetic spectrum (see Fig.
\ref{fig:4}).

To further test the applicability of the effect discussed here on
the physics of the cuprates, it would be desirable to directly measure
the nematic susceptibility in tetragonal cuprates. In analogy to what
has been done for the iron-pnictides (see Ref.~\cite{Fernandes-NatPhys-2014}),
$\chi_{\mathrm{nem}}$ is closely related to several observables,
such as the elastoresistance~\cite{Chu-Science-2012}, the shear modulus, or electronic Raman
scattering. If the biquadratic term found here was to govern the nematic
properties of tetragonal cuprates, such as HgBa$_{2}$CuO$_{4}$,
$\chi_{\mathrm{nem}}$ should be enhanced but not divergent \textendash{}
possibly displaying a peak at a temperature comparable to $J$. Furthermore,
the temperature dependences of $\chi_{\mathrm{nem}}$ in the $B_{1g}$
and $B_{2g}$ channels would be similar, although the former would
be larger. 

Because the biquadratic term is the result of charge fluctuations
on the oxygen $p$-orbitals, it is only present in the hole-doped
side of the phase diagram, since electron-doping adds charge carriers
directly to the Cu sites~\cite{Armitage-RMP-2010}. To the
best of our knowledge, nematic tendencies have not been reported in
electron-doped cuprates~\cite{Kivelson-RMP-2003}. It would be interesting
to verify this effect by experimentally determining $\chi_{\mathrm{nem}}$
in tetragonal electron-doped systems, such as Nd$_{2}$CuO$_{4}$.

\section*{Methods}
\subsection*{Derivation of $t-J-K$-model from microscopic three-band model}
We derive the biquadratic $K$ spin exchange term in Eq.~\eqref{eq:HtJK} from a microscopic interacting three-band model $H = H_0 + H_{U} + H_V$ that takes Cu $d_{x^2-y^2}$ and O $p_x, p_y$ orbitals into account and reads
\begin{align}
H_0 &= \sum_{i, \sigma} \Delta \, n^p_{i\sigma} +  t_{pd} \sum_{\langle i,j \rangle} \bigl( (-1)^{u_{ij}}  d^\dag_{i\sigma} p_{j\sigma} + \text{h.c.} \bigr) \nonumber \\ &+ t_{pp} \sum_{\langle \langle i,j \rangle \rangle} \bigl( (-1)^{u'_{ij}} p^\dag_{i\sigma} p_{j\sigma} + \text{h.c.} \bigr) \\
H_{U}&= U_{dd} \sum_i  n^d_{i\uparrow} n^d_{i \downarrow} + \frac{U_{pp}}{2} \sum_i n^p_{i \uparrow} n^p_{i \downarrow} \\
H_V &= V_{pd} \sum_{\langle i,j \rangle}  n^d_{i} n^p_j + V_{pp} \sum_{\langle \langle i,j \rangle \rangle} n^p_i n^p_j \,.
\label{M-eq:1}
 \end{align}
Here, $d^\dag_{i\sigma}$ creates a hole in the $d_{x^2-y^2}$ orbital at Bravais lattice site $\bfrr_i$, and $p^\dag_{i\sigma}$ creates a hole in the O $p_x$ and $p_y$ orbital $i \equiv \bfrr_i  + \frac{\hat{\bfx}}{2}$ and $i \equiv \bfrr_i + \frac{\hat{\bfy}}{2}$, respectively (see Fig.~\ref{fig:1}(b)). The parameters in the Hamiltonian are the on-site orbital energy difference $\Delta = \epsilon_p - \epsilon_d$, hoppings $t_{pp}$, $t_{pd}$ (see Fig.~\ref{fig:1}), on-site interactions $U_{pp}$, $U_{dd}$ and nearest-neighbor interactions $V_{pp}$, $V_{pd}$. 

As $U_{dd}$ is the largest energy scale, we perform a strong coupling expansion which yields a description in terms of localized Cu-site spins $\bfss_i$ and mobile oxygen holes. The second order terms contain direct O hopping terms and Cu-O Kondo coupling terms. We consider these terms in two complementary limits: (i) assuming an antiferromagnetically ordered background of Cu spins that renormalizes the oxygen bandstructure (main text), and (ii) disregarding both terms which yields a free oxygen dispersion (SI). Both calculations yields qualitatively identical results for quadrupolar response $\tilde{\Pi}^\eta$ (see Fig.~\ref{fig:2}(a)) and biquadratic exchange $K$. To fourth order appears the Heisenberg exchange interaction term and the exchange interaction term in Eq.~\eqref{eq:2} that includes the density of the intermediate O orbital. Upon integrating out the O holes this term yields the biquadratic exchange $K$ term of Eq.~\eqref{eq:HtJK}. 

As the theory is quartic in O hole operators, we must first decouple the interaction terms. We perform the decoupling in the channel of the total and relative density of O atoms in a unit cell: $n^p_i = n^p_{i + \frac{\hat{x}}{2}} + n^p_{i + \frac{\hat{y}}{2}}$ and $\eta^p_i = n^p_{i + \frac{\hat{x}}{2}} - n^p_{i + \frac{\hat{y}}{2}}$. Introducing the vector $\nu_\bfk = (n^p_\bfk, \eta^p_{\bfk})^T$, where $n_i^p = \frac{1}{N_L} \sum_{\bfk} n^p_k e^{i \bfk \cdot \bfrr_i}$, the interaction terms read 
\begin{align}
H_{U_{pp}} + H_{V_{pp}} = -\sum_{\bfk; k_x > 0} \nu_\bfk^\dag U^{-1}_{\bfk} \nu_\bfk
\label{M-eq:2}
\end{align}
with interaction matrix
\begin{align}
U_{\bfk}^{-1} = \frac{2}{N_L} \bmx - \text{Re} \, U_{+,\bfk} & \frac{i V_{pp}}{4} \text{Im} f_\bfk \\ -\frac{i V_{pp}}{4} \text{Im} f_\bfk & \text{Re} \, U_{-, \bfk} \emx \,.
\label{M-eq:3}
\end{align}
Here, $U_\pm = \frac{1}{8} \bigl( 2 V_{pp} f_{\bfk} \pm U_{pp} \bigr)$ and $f_\bfk = 1 + e^{- i k_x} + e^{i k_y} + e^{i (k_y - k_x)}$. We decouple the interactions using a Hubbard-Stratonovich (HS) transformation, which yields an action that is purely quadratic in O operators, but contains the HS fields $\Phi_\bfk = (\psi_\bfk, \phi_\bfk)$:
\begin{align}
S &= - \int_{q, q'} \sum_{\genfrac{}{}{0pt}{}{\alpha, \alpha'}{\sigma, \sigma'} } p^\dag_{Q} G^{-1}_{Q, Q'} p_{Q'} + \int_q \Phi^\dag_q U_{q} \Phi_q\,.
\label{M-eq:4}
\end{align}
Here, $Q = (q, \alpha, \sigma)$ combines the Matsubara frequency-momentum index $q = (i q_n, \bfq)$ with index $\alpha =x,y$, which runs over the O orbital index $p_x, p_y$, and $\sigma$, which denotes spin. The Green's function $G^{-1}$ contains the Cu spin operators due to the coupling term $\propto J'$ in Eq.~\eqref{eq:2}, the O hopping terms and the HS variables: $G^{-1} = G_{0}^{-1} + G^{-1}_{\Phi} + G^{-1}_{J'}$. It is diagonal in spin space, and is in orbital space $(\tau^\alpha)$ given by
\begin{equation}
G_0^{-1}(Q, Q') = (i q_n - \Delta + \mu) \openone - t_{pp} \text{Re}(h_{\bfq}) \tau^x + t_{pp} \text{Im}(h_{\bfq}) \tau^y
\label{M-eq:5}
\end{equation}
with $h_\bfq = 1 - e^{i q_x} - e^{-iq_y} + e^{i (q_x - q_y)}$. Here we suppress the second-order terms of the strong coupling expansion, which simply renormalize the dispersion. If one considers the motion of O holes in the background of AF ordered Cu spins, as we have done to calculate the results of Fig.~\ref{fig:2}(a), the dispersion entering $G_0^{-1}$ is modified accordingly~\cite{Fischer-NJP-2014,SupplMat-CuprateNematicity}. The other terms in the Green's function read
\begin{equation}
[G^{-1}_{\Phi} + G^{-1}_{J'}](Q,Q') = A_{0}(q-q') \tau^0 + A_{z}(q-q') \tau^z \,.
\label{M-eq:5a}
\end{equation}
Here, $\tau^0 \equiv \openone$ and we have defined the two functions 
\begin{align}
A_0(q-q') &= \psi_{q-q'} - \sum_{\bfp} \mathcal{S}_{\bfp, \bfq- \bfq'} h_n(\bfp, \bfq'-\bfq)\\
A_z(q-q') &= \phi_{q-q'} - \sum_{\bfp} \mathcal{S}_{\bfp, \bfq- \bfq'} h_\eta(\bfp, \bfq'-\bfq)
\label{M-eq:5b}
\end{align}
with Cu spin bilinear $\mathcal{S}_{\bfp, \bfq} = \bfss_\bfp \cdot \bfss_{-\bfp - \bfq}$ and lattice functions $h_{n/\eta}(\bfp, \bfk) = \frac{J'}{2} ( e^{i p_x} \pm e^{i p_y} + e^{-i (p_x + k_x)} \pm e^{-i(p_y + k_y)})$, where the upper (lower) sign relates to $h_n$ ($h_\eta$). 

Integrating over the O degrees of freedom results in an action of the form
\begin{equation}
S = \int_q \Phi^\dag_q U_{q} \Phi_q - \text{Tr} \ln (- G^{-1}) \,,
\label{M-eq:6}
\end{equation}
We expand this expression to second order $S_2 = \frac{1}{2} \text{Tr} [ \{ G_0 (G^{-1}_\Phi + G^{-1}_{J'}) \}^2 ]$ in order to find
\begin{equation}
 S_2 = - \frac{1}{2} \int_q \sum_{\alpha, \alpha'} A_{\alpha}(q) A_{\alpha'}(-q) \Pi^{\alpha \alpha'}_{-q} \,,
 \label{M-eq:7}
 \end{equation} 
which includes the biquadratic exchange $K$ term. Here, we have introduced the oxygen density response function
\begin{equation}
\Pi^{\alpha \alpha'}_{q} = - \int_k \text{Tr} \bigl[ G_{0}(k) \tau^\alpha G_{0}(k+q) \tau^{\alpha'} \bigr] \,.
\label{M-eq:8}
\end{equation}
 The bare biquadratic exchange constant $K_0$ is given by the $zz$-component of this response function as
 \begin{equation}
 K_0 = \frac{J'^2}{2} \int_\bfq \Pi^{zz}_\bfq \,.
 \label{M-eq:9}
 \end{equation}
Note that we write $\Pi^{zz}_\bfq \equiv \Pi^\eta_\bfq$ in the main text. It is straightforward to obtain the biquadratic exchange renormalized by O density fluctuations by performing the Gaussian integration over the HS fields in Eq.~\eqref{M-eq:6}, which yields the renormalized response function
\begin{equation}
\tilde{\Pi}^{\alpha \alpha'}_{\bfq} = \Pi^{\alpha \alpha'}_{\bfq} + \frac{1}{2} \sum_{\beta, \beta'} \bigl(\tilde{U}^{-1}_{\bfq} \bigr)_{\beta \beta'} \Pi^{\beta \alpha}_\bfq \Pi^{\beta' \alpha'}_{\bfq}
\label{M-eq:10}
\end{equation}
where we have defined $(\tilde{U}_q) = (U_\bfq)_{\alpha \alpha'} - \frac{1}{2} \Pi^{\alpha \alpha'}_{q}$. Approximating the local response by the long-wavelength $\bfq= 0$ component $\tilde{\Pi}^{zz}_{ii}\approx \tilde{\Pi}^{zz}_{\bfq = 0}$ yields for renormalized biquadratic exchange constant $K$ as given in Eq.~\eqref{eq:5} of the main text. 
\subsection*{Nematic susceptibility within soft-spin quantum field theory}
In the main text, we analyze the nematic susceptibility $\chi_{\text{nem}}$ in the $t$-$J$-$K$-model using a soft-spin quantum field theory. This allows us to investigate the effect of quantum fluctutations on the nematic response. Our main results are shown in Fig.~\ref{fig:3}. After decoupling the biquadratic $K$ term using HS variable $\varphi_r$ the soft-spin action reads
\begin{flalign}
S &= \gamma \int_q \Bigl[ r_0 + q^2 + (\varphi_r + h_\varphi) (q_x^2 - q_y^2) + \gamma |\omega_n|^{\frac{2}{z}} \Bigr] M^\alpha_{q} M^\alpha_{-q} \nonumber \\
& + \frac{\gamma^2 u}{2 N} \int_{q_1, q_2, q_3} M^\alpha_{q_1} M^\alpha_{q_2} M^\beta_{q_3} M^\beta_{-q_1- q_2 - q_3} + \int_r \frac{N \varphi_r}{2 g}&\,.
\label{M-eq:11}
\end{flalign}
Here, $\bfmm_q$ denotes an $N$-component N\'eel magnetization order parameter ($N=3$ in the physical system) and summation over repeated indices $\alpha, \beta$ is implied. The integrations are over $\int_r = \int_0^{1/T}d \tau \int d^2r$ and $\int_q = T \sum_{\omega_n}^{\Lambda_\omega} \int^\Lambda \frac{d^2 \bfq}{(2 \pi)^2}$ up to some dimensionless momentum and frequency cutoffs $\Lambda$ and $\gamma \Lambda_\omega$. The parameter $r_0$ controls the distance to the quantum critical point separating a N\'eel ordered regime from a quantum disordered paramagnetic regime, $u$ is an interaction constant and the coupling constant $g \propto K/J$ is proportional to the biquadratic exchange. We have added a source field $h_\varphi \equiv h_r$ that couples to homogeneous nematic order. We use a dynamic critical exponent of $z=2$ in the following, which describes damping due to particle-hole excitations in the presence of mobile holes. 

The nematic susceptibility in Eq.~\eqref{eq:chi)nem} can be calculated from the partition function $Z = \int \mathcal{D}(\bfmm_q, \varphi_r) e^{-S}$ as 
\begin{equation}
\chi_{\text{nem}} = \frac{T}{L^2} \frac{\partial^2 Z}{\partial h_\varphi^2}\Bigr|_{h_\varphi=0} = \frac{\chi_{\text{nem},0}}{1 - \frac{g}{N} \chi_{\text{nem},0}}
\label{M-eq:12}
\end{equation}
where $L$ is the linear system size and the bare nematic susceptibility is given by 
\begin{equation}
\chi_{\text{nem},0} = \frac{N}{g} - \frac{T}{L^2 \av{\bar{\varphi}^2_r}}\,,
\label{M-eq:13}
\end{equation}
with $\bar{\varphi}_r = \varphi_r + h_\varphi$. In the following, we consider homogeneous HS fields $\varphi_r$, $\bar{\varphi_r}$. To calculate the expectation value $\av{\bar{\varphi}^2_r}$, we first decouple the quartic $u$-term in Eq.~\eqref{M-eq:11} using HS field $\psi$, then separate longitudinal and transverse components $\bfmm_r = (\sqrt{N} M, \boldsymbol{\pi}_r)$ and integrate over the transverse ones to arrive at the (dimensionless) action $s  \equiv S/[L^2 (\gamma T)^{-1}]$ given by 
\begin{flalign}
s & = N r M^2  + \frac{N (\bar{\varphi}_r - h_\varphi)^2}{2 \tilde{g}} - \frac{\psi^2}{2 \tilde{u}} \nonumber \\
& \qquad + \frac{N-1}{2} \gamma \int_q \ln \Bigl( r_q + \bar{\varphi}_r (q_x^2 - q_y^2) \Bigr) &\,.
\label{M-eq:14}
\end{flalign}
Here, $r_q = r + \bfq^2 + \gamma |\omega_n|$, $r = r_0 + \psi$ and we have defined dimensionless interaction constants $\tilde{g} = g/\gamma$ and $\tilde{u} = u/\gamma$. Next, we expand the logarithm in small $\bar{\varphi}_r$ up to second order and obtain Eq.~\eqref{M-eq:13} by differentiation as
\begin{equation}
\chi_{\text{nem},0} = \frac{N T}{2} \sum_{\omega_n} \int^\Lambda \frac{d^2 q}{(2 \pi)^2} \frac{q^4 \cos^2(2 \theta)}{(r + q^2 + \gamma |\omega_n|)^2} \,.
\label{M-eq:15}
\end{equation}
We can exactly perform the summation over Matsubara frequencies, the momentum integration and then absorb the cutoff $\Lambda$ by expressing $\chi_{\text{nem,0}}$ in terms of the dimensionless variables $\tilde{T}=\gamma T/\Lambda^2$ and $\tilde{r} = r/\Lambda^2$. The lengthy expression is given in the Supplemental Material~\cite{SupplMat-CuprateNematicity} together with a three-dimensional plot as a function of $\tilde{T}$ and $\tilde{r}$. Cuts for different functional behaviors of the magnetic N\'eel correlation length on temperature $r(T) \equiv \xi^{-2}(T)$ are shown in Fig.~\ref{fig:3}(a). 

We can derive the functional behavior of $r(T)$ within a large-$N$ approach, where we need to solve the following well-known self-consistency equation 
\begin{equation}
r = r_0 + u M^2 + \frac{u}{2} \int_q \frac{1}{r + q^2 + \gamma |\omega_n|} \,.
\label{M-eq:16}
\end{equation}
Solving this equation requires us to introduce a finite frequency cutoff $\Lambda_\omega$, but the qualitative behavior of $\chi_{\text{nem},0}$ and $\chi_{\text{nem}}$ does not depend on the cutoff choice as long as $\Lambda, \Lambda_\omega \gg r, \gamma T$. The results for $\chi_{\text{nem},0}$ and $\chi_{\text{nem}}$ shown in Fig.~\ref{fig:3}(b, c) are obtained from the large-$N$ solution of $r(T)$ for fixed parameters $\tilde{u}, \Lambda, \Lambda_\omega$ and distance to the quantum-critical point $\delta r_0 = r - r_{0,c}$.

\subsection*{Details on the classical Monte-Carlo simulations}
The Monte Carlo simulations were carried out at $100$ equally spaced temperature points in the interval $0.001<T/J<2.971$. We applied a combination of single-move Metropolis Monte Carlo steps and non-local parallel-tempering-exchange steps between neighboring temperature configurations. The simulations shown in Fig.~\ref{fig:2}(b, c) of the main text were carried out for systems of $40\times 40$ spins and biquadratic exchange couplings $K/J = \{0.0, 0,35, 0.45\}$. We consider a ferromagnetic next-nearest-neighbor exchange coupling $J_2 = -0.1 J$ as well. Note that the ground state phase transition in the classical model between N\'eel and collinear order occurs at $J/2 = J_2 + K$. Following thermalization, the averages were computed for each temperature with at least $4.5\times 10^6$ Monte Carlo sweeps (MCS). The error bars were estimated by using the well-known Jackknife procedure.

Finally, we mention that we have performed Monte-Carlo simulations also for the purely bilinear spin Hamiltonian that is obtained from $H_{J-K}$ by using the well-known relations valid for spin-$1/2$ operators: $\bigl( \bfss_i\cdot\bfss_j \bigr)^2 = \frac{3}{16} - \frac12 \bfss_i \cdot \bfss_j$ and $\bigl( \bfss_i \cdot \bfss_j \bigr) \bigl( \bfss_i \cdot \bfss_k \bigr) = \frac14 \bfss_j \cdot \bfss_k + \frac{i}{2} \bfss_i \cdot (\bfss_j \times \bfss_k)$. These allow rewriting the biquadratic $K$ term as a sum of three \emph{bilinear} spin exchange terms 
\begin{flalign}
  \label{eq:49}
  &\tilde{H}_{J-K} = \frac12 \bigl( J + \frac{K}{4 S^2} \bigr) \sum_i \sum_\delta \bfss_i \cdot \bfss_{i + \delta}  \\ & \quad + \frac{K}{8 S^2} \sum_{i} \sum_{\delta'} \bfss_{i} \cdot \bfss_{i + \delta'} 
  - \frac{K}{16 S^2} \sum_i \sum_{\delta''} \bfss_i \cdot \bfss_{i + \delta''} \nonumber\,.& 
\end{flalign}
Here, $\delta (\delta')$ runs over the (next-)nearest neighbors of the square lattice and $\delta''$ runs over the second-neighbors along the bonds. Importantly, classical Monte-Carlo simulation results for this Hamiltonian $\tilde{H}_{J-K}$ show the same enhancement of the nematic susceptibility $\chi_{\text{nem}}$ as a function of $K$ as results for the original Hamiltonian $H_{J-K}$ that includes the biquadratic exchange term. 

\section*{Acknowledgments}

We gratefully acknowledge helpful discussions with A.~V.~Chubukov, M.-H.
Julien, B.~Keimer, M.~Le~Tacon, and L. Taillefer. 

\section*{Competing interests}

The authors declare no competing interests.

\section*{Author contributions}

P.P.O., B.J., R.M.F. and J.S. contributed extensively to the calculations,
prepared the figures and wrote the paper.

\section*{Funding}
P.P.O. acknowledges support from Iowa State University Startup Funds. J.S. acknowledges
financial support by the Deutsche Forschungsgemeinschaft through Grant
No.~SCHM 1031/7-1. This work was carried out using the computational
resource bwUniCluster funded by the Ministry of Science, Research
and Arts and the Universities of the State of Baden-W\"urttemberg, Germany,
within the framework program bwHPC.

\section*{Data Availability}
The data that support the findings of this study are available from the authors upon request.



\begin{thebibliography}{10}
\expandafter\ifx\csname url\endcsname\relax
  \def\url#1{\texttt{#1}}\fi
\expandafter\ifx\csname urlprefix\endcsname\relax\def\urlprefix{URL }\fi
\providecommand{\bibinfo}[2]{#2}
\providecommand{\eprint}[2][]{\url{#2}}

\bibitem{Wu-Nature-2011}
\bibinfo{author}{Wu, T.} \emph{et~al.}
\newblock \bibinfo{title}{Magnetic-field-induced charge-stripe order in the
  high-temperature superconductor {Y}{Ba}$_2${Cu}$_3${O}$_y$}.
\newblock \emph{\bibinfo{journal}{Nature}} \textbf{\bibinfo{volume}{477}},
  \bibinfo{pages}{191--194} (\bibinfo{year}{2011}).

\bibitem{Ghiringhelli-2012}
\bibinfo{author}{Ghiringhelli, G.} \emph{et~al.}
\newblock \bibinfo{title}{Long-range incommensurate charge fluctuations in
  ({Y}, {Nd}){Ba}$_2${Cu}$_3${O}$_{6+x}$}.
\newblock \emph{\bibinfo{journal}{Science}} \textbf{\bibinfo{volume}{337}},
  \bibinfo{pages}{821--825} (\bibinfo{year}{2012}).

\bibitem{Chang2012}
\bibinfo{author}{Chang, J.} \emph{et~al.}
\newblock \bibinfo{title}{Direct observation of competition between
  superconductivity and charge density wave order in
  {YBa}$_2${Cu}$_3${O}$_{6.67}$}.
\newblock \emph{\bibinfo{journal}{Nat. Phys.}} \textbf{\bibinfo{volume}{8}},
  \bibinfo{pages}{871--876} (\bibinfo{year}{2012}).

\bibitem{LeBoeuf2012}
\bibinfo{author}{LeBoeuf, D.} \emph{et~al.}
\newblock \bibinfo{title}{Thermodynamic phase diagram of static charge order in
  underdoped {YBa}$_2${Cu}$_3${O}$_y$}.
\newblock \emph{\bibinfo{journal}{Nat. Phys.}} \textbf{\bibinfo{volume}{9}},
  \bibinfo{pages}{79--83} (\bibinfo{year}{2012}).

\bibitem{Ando2002}
\bibinfo{author}{Ando, Y.}, \bibinfo{author}{Segawa, K.},
  \bibinfo{author}{Komiya, S.} \& \bibinfo{author}{Lavrov, A.~N.}
\newblock \bibinfo{title}{Electrical resistivity anisotropy from self-organized
  one dimensionality in high-temperature superconductors}.
\newblock \emph{\bibinfo{journal}{Phys. Rev. Lett.}}
  \textbf{\bibinfo{volume}{88}}, \bibinfo{pages}{137005}
  (\bibinfo{year}{2002}).

\bibitem{Hinkov-Science-2008}
\bibinfo{author}{Hinkov, V.} \emph{et~al.}
\newblock \bibinfo{title}{Electronic liquid crystal state in the
  high-temperature superconductor {YBa}$_2${Cu}$_3${O}$_{6.45}$}.
\newblock \emph{\bibinfo{journal}{Science}} \textbf{\bibinfo{volume}{319}},
  \bibinfo{pages}{597--600} (\bibinfo{year}{2008}).

\bibitem{Daou-Nature-2010}
\bibinfo{author}{Daou, R.} \emph{et~al.}
\newblock \bibinfo{title}{Broken rotational symmetry in the pseudogap phase of
  a high-{Tc} superconductor}.
\newblock \emph{\bibinfo{journal}{Nature}} \textbf{\bibinfo{volume}{463}},
  \bibinfo{pages}{519--522} (\bibinfo{year}{2010}).

\bibitem{Lawler-Nature-2010}
\bibinfo{author}{Lawler, M.~J.} \emph{et~al.}
\newblock \bibinfo{title}{Intra-unit-cell electronic nematicity of the
  high-{Tc} copper-oxide pseudogap states}.
\newblock \emph{\bibinfo{journal}{Nature}} \textbf{\bibinfo{volume}{466}},
  \bibinfo{pages}{347--351} (\bibinfo{year}{2010}).

\bibitem{Cyr-PRB-2015}
\bibinfo{author}{Cyr-Choini\`ere, O.} \emph{et~al.}
\newblock \bibinfo{title}{Two types of nematicity in the phase diagram of the
  cuprate superconductor {YBa}$_2${Cu}$_3${O}$_y$}.
\newblock \emph{\bibinfo{journal}{Phys. Rev. B}} \textbf{\bibinfo{volume}{92}},
  \bibinfo{pages}{224502} (\bibinfo{year}{2015}).

\bibitem{Ramshaw-npj-2017}
\bibinfo{author}{Ramshaw, B.~J.} \emph{et~al.}
\newblock \bibinfo{title}{Broken rotational symmetry on the {F}ermi surface of
  a high-{Tc} superconductor}.
\newblock \emph{\bibinfo{journal}{npj Quantum Mater.}}
  \textbf{\bibinfo{volume}{2}} (\bibinfo{year}{2017}).

\bibitem{Kivelson-RMP-2003}
\bibinfo{author}{Kivelson, S.~A.} \emph{et~al.}
\newblock \bibinfo{title}{How to detect fluctuating stripes in the
  high-temperature superconductors}.
\newblock \emph{\bibinfo{journal}{Rev. Mod. Phys.}}
  \textbf{\bibinfo{volume}{75}}, \bibinfo{pages}{1201--1241}
  (\bibinfo{year}{2003}).

\bibitem{Vojta-AdvPhys-2009}
\bibinfo{author}{Vojta, M.}
\newblock \bibinfo{title}{Lattice symmetry breaking in cuprate superconductors:
  stripes, nematics, and superconductivity}.
\newblock \emph{\bibinfo{journal}{Adv. Phys.}} \textbf{\bibinfo{volume}{58}},
  \bibinfo{pages}{699--820} (\bibinfo{year}{2009}).

\bibitem{KeimerKivelson-Nature-2015}
\bibinfo{author}{Keimer, B.}, \bibinfo{author}{Kivelson, S.~A.},
  \bibinfo{author}{Norman, M.~R.}, \bibinfo{author}{Uchida, S.} \&
  \bibinfo{author}{Zaanen, J.}
\newblock \bibinfo{title}{From quantum matter to high-temperature
  superconductivity in copper oxides}.
\newblock \emph{\bibinfo{journal}{Nature}} \textbf{\bibinfo{volume}{518}},
  \bibinfo{pages}{179--186} (\bibinfo{year}{2015}).

\bibitem{Fradkin2015}
\bibinfo{author}{Fradkin, E.}, \bibinfo{author}{Kivelson, S.~A.} \&
  \bibinfo{author}{Tranquada, J.~M.}
\newblock \bibinfo{title}{Colloquium: Theory of intertwined orders in high
  temperature superconductors}.
\newblock \emph{\bibinfo{journal}{Rev. Mod. Phys.}}
  \textbf{\bibinfo{volume}{87}}, \bibinfo{pages}{457--482}
  (\bibinfo{year}{2015}).

\bibitem{Kivelson-Nature-98}
\bibinfo{author}{Kivelson, S.~A.}, \bibinfo{author}{Fradkin, E.} \&
  \bibinfo{author}{Emery, V.~J.}
\newblock \bibinfo{title}{Electronic liquid-crystal phases of a doped mott
  insulator}.
\newblock \emph{\bibinfo{journal}{Nature}} \textbf{\bibinfo{volume}{393}},
  \bibinfo{pages}{550} (\bibinfo{year}{1998}).

\bibitem{Yamase-JPhysSocJpn-2000}
\bibinfo{author}{Yamase, H.} \& \bibinfo{author}{Kohno, H.}
\newblock \bibinfo{title}{Instability toward formation of quasi-one-dimensional
  {F}ermi surface in two-dimensional {t-J} model}.
\newblock \emph{\bibinfo{journal}{J. Phys. Soc. Jpn.}}
  \textbf{\bibinfo{volume}{69}}, \bibinfo{pages}{2151} (\bibinfo{year}{2000}).

\bibitem{Kivelson-PRB-2004}
\bibinfo{author}{Kivelson, S.~A.}, \bibinfo{author}{Fradkin, E.} \&
  \bibinfo{author}{Geballe, T.~H.}
\newblock \bibinfo{title}{Quasi-one-dimensional dynamics and nematic phases in
  the two-dimensional emery model}.
\newblock \emph{\bibinfo{journal}{Phys. Rev. B}} \textbf{\bibinfo{volume}{69}},
  \bibinfo{pages}{144505} (\bibinfo{year}{2004}).

\bibitem{Yamase-PRB-2006}
\bibinfo{author}{Yamase, H.} \& \bibinfo{author}{Metzner, W.}
\newblock \bibinfo{title}{Magnetic excitations and their anisotropy in
  {YBa}$_{2}${Cu}$_{3}${O}$_{6+x}$: Slave-boson mean-field analysis of the
  bilayer {t-J} model}.
\newblock \emph{\bibinfo{journal}{Phys. Rev. B}} \textbf{\bibinfo{volume}{73}},
  \bibinfo{pages}{214517} (\bibinfo{year}{2006}).

\bibitem{Yamase-PRB-2009}
\bibinfo{author}{Yamase, H.}
\newblock \bibinfo{title}{Theory of reduced singlet pairing without the
  underlying state of charge stripes in the high-temperature superconductor
  {Y}{Ba}$_2${Cu}$_3${O}$_{6.45}$}.
\newblock \emph{\bibinfo{journal}{Phys. Rev. B}} \textbf{\bibinfo{volume}{79}},
  \bibinfo{pages}{052501} (\bibinfo{year}{2009}).

\bibitem{Okamoto-PRB-2010}
\bibinfo{author}{Okamoto, S.}, \bibinfo{author}{S\'en\'echal, D.},
  \bibinfo{author}{Civelli, M.} \& \bibinfo{author}{Tremblay, A.-M.~S.}
\newblock \bibinfo{title}{Dynamical electronic nematicity from {M}ott physics}.
\newblock \emph{\bibinfo{journal}{Phys. Rev. B}} \textbf{\bibinfo{volume}{82}},
  \bibinfo{pages}{180511} (\bibinfo{year}{2010}).

\bibitem{Fischer-PRB-2011}
\bibinfo{author}{Fischer, M.~H.} \& \bibinfo{author}{Kim, E.-A.}
\newblock \bibinfo{title}{Mean-field analysis of intra-unit-cell order in the
  emery model of the {CuO}${}_{2}$ plane}.
\newblock \emph{\bibinfo{journal}{Phys. Rev. B}} \textbf{\bibinfo{volume}{84}},
  \bibinfo{pages}{144502} (\bibinfo{year}{2011}).

\bibitem{Andersen-EPL-2012}
\bibinfo{author}{Andersen, B.~M.}, \bibinfo{author}{Graser, S.} \&
  \bibinfo{author}{Hirschfeld, P.~J.}
\newblock \bibinfo{title}{Correlation and disorder-enhanced nematic spin
  response in superconductors with weakly broken rotational symmetry}.
\newblock \emph{\bibinfo{journal}{Europhys. Lett.}}
  \textbf{\bibinfo{volume}{97}}, \bibinfo{pages}{47002} (\bibinfo{year}{2012}).

\bibitem{Bulut-PRB-2013}
\bibinfo{author}{Bulut, S.}, \bibinfo{author}{Atkinson, W.~A.} \&
  \bibinfo{author}{Kampf, A.~P.}
\newblock \bibinfo{title}{Spatially modulated electronic nematicity in the
  three-band model of cuprate superconductors}.
\newblock \emph{\bibinfo{journal}{Phys. Rev. B}} \textbf{\bibinfo{volume}{88}},
  \bibinfo{pages}{155132} (\bibinfo{year}{2013}).

\bibitem{Fischer-NJP-2014}
\bibinfo{author}{Fischer, M.~H.}, \bibinfo{author}{Wu, S.},
  \bibinfo{author}{Lawler, M.}, \bibinfo{author}{Paramekanti, A.} \&
  \bibinfo{author}{Kim, E.-A.}
\newblock \bibinfo{title}{Nematic and spin-charge orders driven by hole-doping
  a charge-transfer insulator}.
\newblock \emph{\bibinfo{journal}{New J. Phys.}} \textbf{\bibinfo{volume}{16}},
  \bibinfo{pages}{093057} (\bibinfo{year}{2014}).

\bibitem{Volkov-PRB-2016}
\bibinfo{author}{Volkov, P.~A.} \& \bibinfo{author}{Efetov, K.~B.}
\newblock \bibinfo{title}{Spin-fermion model with overlapping hot spots and
  charge modulation in cuprates}.
\newblock \emph{\bibinfo{journal}{Phys. Rev. B}} \textbf{\bibinfo{volume}{93}},
  \bibinfo{pages}{085131} (\bibinfo{year}{2016}).

\bibitem{Chubukov-PRB-2014}
\bibinfo{author}{Wang, Y.} \& \bibinfo{author}{Chubukov, A.}
\newblock \bibinfo{title}{Charge-density-wave order with momentum $(2q,0)$ and
  $(0,2q)$ within the spin-fermion model: Continuous and discrete symmetry
  breaking, preemptive composite order, and relation to pseudogap in hole-doped
  cuprates}.
\newblock \emph{\bibinfo{journal}{Phys. Rev. B}} \textbf{\bibinfo{volume}{90}},
  \bibinfo{pages}{035149} (\bibinfo{year}{2014}).

\bibitem{Schuett-PRL-2015}
\bibinfo{author}{Sch\"utt, M.} \& \bibinfo{author}{Fernandes, R.~M.}
\newblock \bibinfo{title}{Antagonistic in-plane resistivity anisotropies from
  competing fluctuations in underdoped cuprates}.
\newblock \emph{\bibinfo{journal}{Phys. Rev. Lett.}}
  \textbf{\bibinfo{volume}{115}}, \bibinfo{pages}{027005}
  (\bibinfo{year}{2015}).

\bibitem{Nie-PRB-2017}
\bibinfo{author}{Nie, L.}, \bibinfo{author}{Maharaj, A.~V.},
  \bibinfo{author}{Fradkin, E.} \& \bibinfo{author}{Kivelson, S.~A.}
\newblock \bibinfo{title}{Vestigial nematicity from spin and/or charge order in
  the cuprates}.
\newblock \emph{\bibinfo{journal}{Phys. Rev. B}} \textbf{\bibinfo{volume}{96}},
  \bibinfo{pages}{085142} (\bibinfo{year}{2017}).

\bibitem{Scheurer-PRL-17}
\bibinfo{author}{Chatterjee, S.}, \bibinfo{author}{Sachdev, S.} \&
  \bibinfo{author}{Scheurer, M.~S.}
\newblock \bibinfo{title}{Intertwining topological order and broken symmetry in
  a theory of fluctuating spin-density waves}.
\newblock \emph{\bibinfo{journal}{Phys. Rev. Lett.}}
  \textbf{\bibinfo{volume}{119}}, \bibinfo{pages}{227002}
  (\bibinfo{year}{2017}).

\bibitem{Tsuchiizu-PRB-2018}
\bibinfo{author}{Tsuchiizu, M.}, \bibinfo{author}{Kawaguchi, K.},
  \bibinfo{author}{Yamakawa, Y.} \& \bibinfo{author}{Kontani, H.}
\newblock \bibinfo{title}{Multistage electronic nematic transitions in cuprate
  superconductors: A functional-renormalization-group analysis}.
\newblock \emph{\bibinfo{journal}{Phys. Rev. B}} \textbf{\bibinfo{volume}{97}},
  \bibinfo{pages}{165131} (\bibinfo{year}{2018}).

\bibitem{Zaanen-PRL-1985}
\bibinfo{author}{Zaanen, J.}, \bibinfo{author}{Sawatzky, G.~A.} \&
  \bibinfo{author}{Allen, J.~W.}
\newblock \bibinfo{title}{Band gaps and electronic structure of
  transition-metal compounds}.
\newblock \emph{\bibinfo{journal}{Phys. Rev. Lett.}}
  \textbf{\bibinfo{volume}{55}}, \bibinfo{pages}{418--421}
  (\bibinfo{year}{1985}).

\bibitem{Emery-PRL-1987}
\bibinfo{author}{Emery, V.~J.}
\newblock \bibinfo{title}{Theory of high-{T}$_{\mathrm{c}}$ superconductivity
  in oxides}.
\newblock \emph{\bibinfo{journal}{Phys. Rev. Lett.}}
  \textbf{\bibinfo{volume}{58}}, \bibinfo{pages}{2794} (\bibinfo{year}{1987}).

\bibitem{Lee-RMP-2006}
\bibinfo{author}{{L}ee, P.~A.}, \bibinfo{author}{{N}agaosa, N.} \&
  \bibinfo{author}{{W}en, X.-G.}
\newblock \bibinfo{title}{{D}oping a {M}ott insulator: {P}hysics of
  high-temperature superconductivity}.
\newblock \emph{\bibinfo{journal}{Rev. Mod. Phys.}}
  \textbf{\bibinfo{volume}{78}}, \bibinfo{pages}{17} (\bibinfo{year}{2006}).

\bibitem{Zaanen-PRB-1988}
\bibinfo{author}{Zaanen, J.} \& \bibinfo{author}{Ole\ifmmode~\acute{s}\else
  \'{s}\fi{}, A.~M.}
\newblock \bibinfo{title}{Canonical perturbation theory and the two-band model
  for high-${T}_{c}$ superconductors}.
\newblock \emph{\bibinfo{journal}{Phys. Rev. B}} \textbf{\bibinfo{volume}{37}},
  \bibinfo{pages}{9423} (\bibinfo{year}{1988}).

\bibitem{Kolley-JPhysC-1992}
\bibinfo{author}{Kolley, E.}, \bibinfo{author}{Kolley, W.} \&
  \bibinfo{author}{Tiertz, R.}
\newblock \bibinfo{title}{Fourth-order interactions in the canonically
  transformed d-p model for {Cu-O} superconductors}.
\newblock \emph{\bibinfo{journal}{J. Phys. C}} \textbf{\bibinfo{volume}{4}},
  \bibinfo{pages}{3517} (\bibinfo{year}{1992}).

\bibitem{ZhangRice-PRB-1988}
\bibinfo{author}{Zhang, F.~C.} \& \bibinfo{author}{Rice, T.~M.}
\newblock \bibinfo{title}{Effective {H}amiltonian for the superconducting {C}u
  oxides}.
\newblock \emph{\bibinfo{journal}{Phys. Rev. B}} \textbf{\bibinfo{volume}{37}},
  \bibinfo{pages}{3759} (\bibinfo{year}{1988}).

\bibitem{ChakravartyHalperinNelson-PRB-1989}
\bibinfo{author}{{C}hakravarty, S.}, \bibinfo{author}{{H}alperin, B.~I.} \&
  \bibinfo{author}{{N}elson, D.~R.}
\newblock \bibinfo{title}{{T}wo-dimensional quantum {H}eisenberg
  antiferromagnet at low temperatures}.
\newblock \emph{\bibinfo{journal}{Phys. Rev. B}} \textbf{\bibinfo{volume}{39}},
  \bibinfo{pages}{2344--2371} (\bibinfo{year}{1989}).

\bibitem{Chubukov-PRB-1994}
\bibinfo{author}{Chubukov, A.~V.}, \bibinfo{author}{Sachdev, S.} \&
  \bibinfo{author}{Ye, J.}
\newblock \bibinfo{title}{Theory of two-dimensional quantum heisenberg
  antiferromagnets with a nearly critical ground state}.
\newblock \emph{\bibinfo{journal}{Phys. Rev. B}} \textbf{\bibinfo{volume}{49}},
  \bibinfo{pages}{11919--11961} (\bibinfo{year}{1994}).

\bibitem{Fernandes-PRB-2012}
\bibinfo{author}{{F}ernandes, R.~M.}, \bibinfo{author}{{C}hubukov, A.~V.},
  \bibinfo{author}{{K}nolle, J.}, \bibinfo{author}{{E}remin, I.} \&
  \bibinfo{author}{{S}chmalian, J.}
\newblock \bibinfo{title}{{P}reemptive nematic order, pseudogap, and orbital
  order in the iron pnictides}.
\newblock \emph{\bibinfo{journal}{Phys. Rev. B}} \textbf{\bibinfo{volume}{85}},
  \bibinfo{pages}{024534} (\bibinfo{year}{2012}).

\bibitem{Shraiman-PRL-1988}
\bibinfo{author}{Shraiman, B.~I.} \& \bibinfo{author}{Siggia, E.~D.}
\newblock \bibinfo{title}{Mobile vacancies in a quantum {H}eisenberg
  antiferromagnet}.
\newblock \emph{\bibinfo{journal}{Phys. Rev. Lett.}}
  \textbf{\bibinfo{volume}{61}}, \bibinfo{pages}{467--470}
  (\bibinfo{year}{1988}).

\bibitem{Sushkov-PRL-2005}
\bibinfo{author}{Sushkov, O.~P.} \& \bibinfo{author}{Kotov, V.~N.}
\newblock \bibinfo{title}{Theory of incommensurate magnetic correlations across
  the insulator-superconductor transition of underdoped
  {La}$_{2-x}${Sr}$_x${CuO}$_{4}$}.
\newblock \emph{\bibinfo{journal}{Phys. Rev. Lett.}}
  \textbf{\bibinfo{volume}{94}}, \bibinfo{pages}{097005}
  (\bibinfo{year}{2005}).

\bibitem{Gabay-PhysicaC-1989}
\bibinfo{author}{Gabay, M.} \& \bibinfo{author}{Hirschfeld, P.}
\newblock \bibinfo{title}{Incommensurate magnetic phases in doped high tc
  compounds}.
\newblock \emph{\bibinfo{journal}{Physica C}}
  \textbf{\bibinfo{volume}{162-164}}, \bibinfo{pages}{823--824}
  (\bibinfo{year}{1989}).

\bibitem{Nafradi-PRL-2016}
\bibinfo{author}{N\'afr\'adi, B.} \emph{et~al.}
\newblock \bibinfo{title}{Magnetostriction and magnetostructural domains in
  antiferromagnetic {YBa}$_2${Cu}$_3${O}$_6$}.
\newblock \emph{\bibinfo{journal}{Phys. Rev. Lett.}}
  \textbf{\bibinfo{volume}{116}}, \bibinfo{pages}{047001}
  (\bibinfo{year}{2016}).

\bibitem{Coldea-PRL-2001}
\bibinfo{author}{Coldea, R.} \emph{et~al.}
\newblock \bibinfo{title}{Spin waves and electronic interactions in
  {La}$_2${CuO}$_4$}.
\newblock \emph{\bibinfo{journal}{Phys. Rev. Lett.}}
  \textbf{\bibinfo{volume}{86}}, \bibinfo{pages}{5377--5380}
  (\bibinfo{year}{2001}).

\bibitem{Haug-NJP-2010}
\bibinfo{author}{Haug, D.} \emph{et~al.}
\newblock \bibinfo{title}{Neutron scattering study of the magnetic phase
  diagram of underdoped {YBa}$_2${Cu}$_3${O}$_{6+ x}$}.
\newblock \emph{\bibinfo{journal}{New J. Phys.}} \textbf{\bibinfo{volume}{12}},
  \bibinfo{pages}{105006} (\bibinfo{year}{2010}).

\bibitem{Fernandes-NatPhys-2014}
\bibinfo{author}{Fernandes, R.~M.}, \bibinfo{author}{Chubukov, A.~V.} \&
  \bibinfo{author}{Schmalian, J.}
\newblock \bibinfo{title}{What drives nematic order in
  iron-based~superconductors?}
\newblock \emph{\bibinfo{journal}{Nat. Phys.}} \textbf{\bibinfo{volume}{10}},
  \bibinfo{pages}{97--104} (\bibinfo{year}{2014}).

\bibitem{Chu-Science-2012}
\bibinfo{author}{Chu, J.-H.}, \bibinfo{author}{Kuo, H.-H.},
  \bibinfo{author}{Analytis, J.~G.} \& \bibinfo{author}{Fisher, I.~R.}
\newblock \bibinfo{title}{Divergent nematic susceptibility in an iron arsenide
  superconductor}.
\newblock \emph{\bibinfo{journal}{Science}} \textbf{\bibinfo{volume}{337}},
  \bibinfo{pages}{710--712} (\bibinfo{year}{2012}).

\bibitem{Armitage-RMP-2010}
\bibinfo{author}{Armitage, N.~P.}, \bibinfo{author}{Fournier, P.} \&
  \bibinfo{author}{Greene, R.~L.}
\newblock \bibinfo{title}{Progress and perspectives on electron-doped
  cuprates}.
\newblock \emph{\bibinfo{journal}{Rev. Mod. Phys.}}
  \textbf{\bibinfo{volume}{82}}, \bibinfo{pages}{2421--2487}
  (\bibinfo{year}{2010}).

\bibitem{SupplMat-CuprateNematicity}
\bibinfo{note}{The Supplemental Material contains details on analytical and
  numerical derivations.}

\end{thebibliography}

\newpage
\pagebreak
\includepdf[pages={{},1,{},2,{},3,{},4,{},5,{},6,{},7,{},8,{},9,{},10,{},11,{},12,{},13,{},14}]{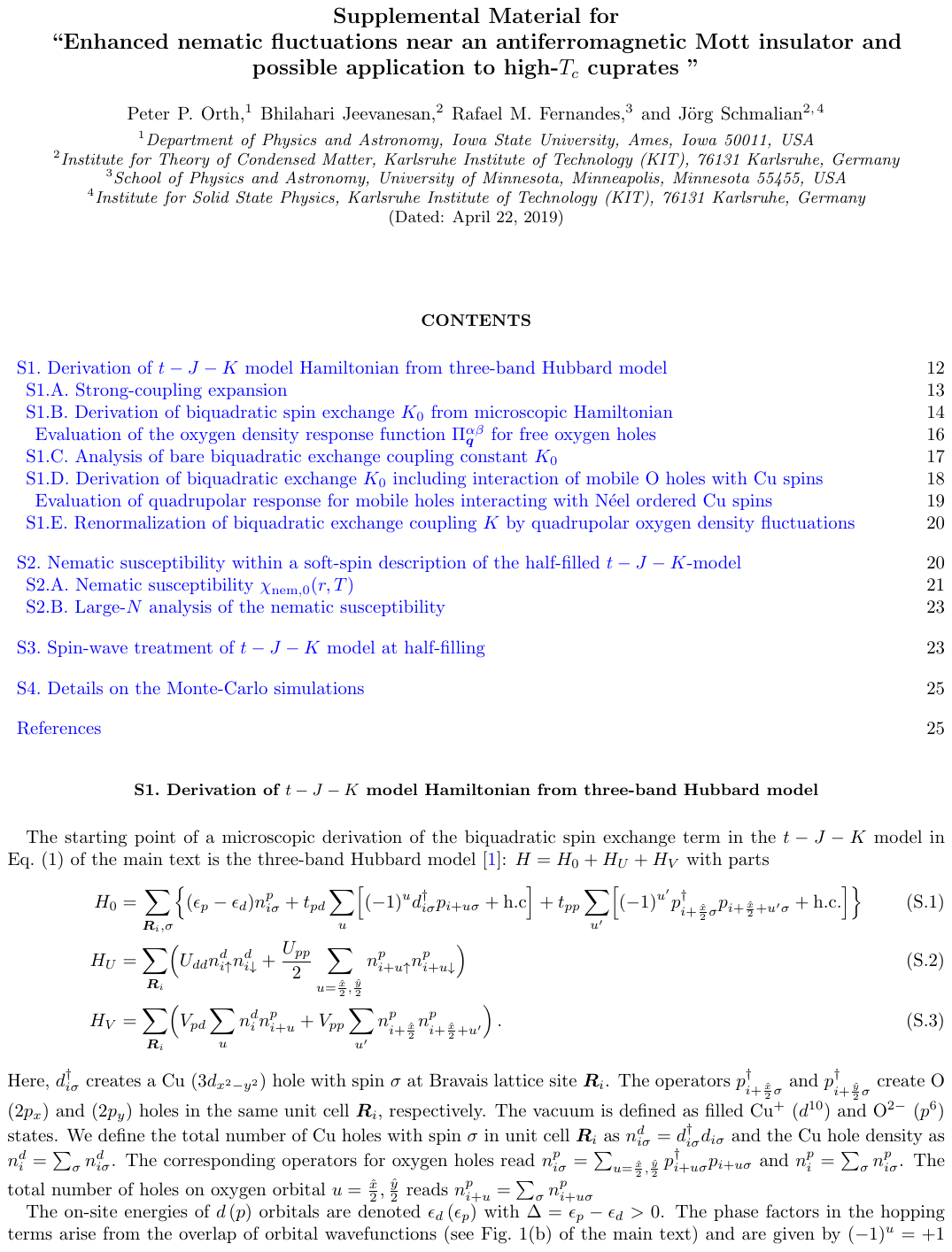}

\end{document}